\pdfoutput=3
\documentclass[aip,amsmath,amssymb]{revtex4-1}

\usepackage{graphicx}%
\usepackage{subcaption}
\usepackage[font=small,labelfont=bf,justification=justified,format=plain]{caption}
\usepackage{multirow}
\usepackage{dcolumn}%
\usepackage{bm}%

\usepackage[utf8]{inputenc}
\usepackage[T1]{fontenc}
\usepackage{mathptmx}
\usepackage{etoolbox}
\usepackage{booktabs}

\makeatletter
\def\@email#1#2{%
 \endgroup
 \patchcmd{\titleblock@produce}
  {\frontmatter@RRAPformat}
  {\frontmatter@RRAPformat{\produce@RRAP{*#1\href{mailto:#2}{#2}}}\frontmatter@RRAPformat}
  {}{}
}%
\makeatother

\usepackage{mathtools}
\newcommand{\nobarfrac}{\genfrac{}{}{0pt}{}}

\begin{document}

\Large
\begin{center}
\texttt{MDRefine}: a Python package for refining Molecular Dynamics trajectories with experimental data

\hspace{10pt}

\large
Ivan Gilardoni$^1$, Valerio Piomponi$^2$, Thorben Fröhlking$^3$, and Giovanni Bussi$^1$ \\

\hspace{10pt}

\small
$^1$ Scuola Internazionale Superiore di Studi Avanzati, SISSA, via Bonomea 265, 34136
Trieste, Italy \\
$^2$ Area Science Park, località Padriciano, 99, 34149 Trieste, Italy \\
$^3$ University of Geneva, 1206 Genève, Switzerland \\
igilardo@sissa.it, bussi@sissa.it\\


\end{center}

\hspace{10pt}


\normalsize

\section*{Abstract}

Molecular dynamics (MD) simulations play a crucial role in resolving the underlying conformational dynamics of molecular systems. However, their capability to correctly reproduce and predict dynamics in agreement with experiments is limited by the accuracy of the force-field model. This capability can be improved by refining the structural ensembles or the force-field parameters. Furthermore, discrepancies with experimental data can be due to imprecise forward models, namely, functions mapping simulated structures to experimental observables.
Here, we introduce \texttt{MDRefine}, a Python package aimed at implementing the refinement of the ensemble, the force field and/or the forward model by comparing MD-generated trajectories with experimental data. The software consists of several tools that can be employed separately from each other or combined together in different ways, providing a seamless interpolation between these three different types of refinement. We use some benchmark cases to show that the combined approach is superior to separately applied refinements. Source code, documentation and examples are freely available at \url{https://pypi.org/project/MDRefine} and \url{https://github.com/bussilab/MDRefine}.

\section*{Introduction}

Molecular dynamics (MD) simulations are a powerful tool to investigate the structure and dynamics of complex molecular systems \cite{hollingsworth2018molecular}.
MD simulations can be used in a predicting fashion. However, due to imperfect sampling and issues
with the employed force fields, their predictions might be incorrect. 
Integration of experimental data is a widely recognized solution to this issue \cite{bottaro2018biophysical} and can be interpreted in two ways:
on one hand, it makes simulations accurate, by forcing them to match experiments.
On the other hand, it makes experiments interpretable, providing a microscopic picture corresponding to the macroscopic
observations and, ultimately, allowing MD simulations to be used as a computational microscope \cite{lee2009discovery}.
Integration of experimental data and MD simulations has been widely exploited in the literature,
in particular for complex molecular and biomolecular systems
(see Refs.~\cite{bonomi2017principles,orioli2020learn,voelz2021reconciling,bernetti2023integrating}
for recent reviews).
Among integrative methods, ensemble refinement techniques where the experimental information is included \emph{a posteriori}
are particularly convenient because they can be used to analyze existing trajectories.
Many methods of this class take advantage of the
maximum entropy principle in order to minimally perturbe the original ensemble while at the same time getting the best possible agreement with experimental values \cite{jaynes1957information,pitera2012use}.
Importantly, measurement devices are always affected by inescapable uncertainties. This uncertainty can be modeled in a Bayesian manner, leading in fact to a regularized version of the maximum entropy approach \cite{hummer2015bayesian,cesari2016combining,cesari2018using,kofinger2019efficient,bottaro2020integrating,borthakur2024towards}, which allows to fine-tune the relative confidence of experiments and prior ensembles by means of a suitably chosen hyperparameter.

When ensemble corrections are only regularized by the requirement of maximizing the relative entropy to the initial prior,
the result
is system-specific and cannot be transferred to different molecular systems.
To restore transferability, the refinement can be performed on the force-field parameters rather than on the ensemble
\cite{norgaard2008experimental,frohlking2020toward}.
This takes place at the expense of a reduced flexibility in the ensemble, which requires chemically motivated insights into the design of the force-field correction. In addition, transferability requires simultaneous analysis of multiple systems,
which in turn complicates the implementation.
Importantly, the maximum entropy principle can be used as a guideline also in force-field fitting strategies, namely by using the relative entropy to regularize the fitting \cite{kofinger2021empirical}.
Recently, the possibility to simultaneously perform ensemble and force-field refinement has been proposed \cite{gilardoni2024boosting}. With a similar formalism, the equations used to connect microscopic ensembles with experimental measurements, so-called \emph{forward models}, can be refined simultaneously with ensemble refinement \cite{frohlking2023simultaneous}.
The generality of the process of adjusting force-field parameters including multiple systems, and the
possibility to combine force-field, ensemble, and forward model refinement,
result in a significant increase in complexity of these methods,
which often require substantial amount of programming.
As far as we know,
general purpose scripts have been published to perform plain ensemble refinement
(see, \emph{e.g.}, \url{https://github.com/KULL-Centre/BME} and \url{https://github.com/bio-phys/BioEn}).
For all the other methods discussed above, a user is forced to write analysis scripts from scratch.

Here we present \texttt{MDRefine}, a Python toolkit to analyse and compare MD-generated trajectories with experimental data. The library contains routines to calculate various quantities (such as relative entropies and $\chi^2$) and to reweight
existing trajectories using different criteria. Importantly, it allows
the generalizations discussed above to be used without the need to implement complex algorithms.
Additionally, it includes an algorithm to determine optimal hyperparameters minimizing the cross-validation error by performing
an optimization procedure rather than an exhaustive hyperparameter scan.
The package is available at \url{https://pypi.org/project/MDRefine}. \texttt{MDRefine} builds on powerful Python libraries such as Numpy \cite{harris2020array}, Scipy \cite{virtanen2020scipy},
and JAX \cite{jax2018github}. It allows to define customized force-field corrections and/or forward models, which best suit for any specific case of interest. The seamless combination of ensembles, force-field and forward-model refinements enables the user both to preliminarily select the desired improvement and/or to determine the optimal one with the given input data.
In this article, we first introduce the basic concept and design of \texttt{MDRefine} and then show how to use it in some benchmark cases. Source code, documentation and illustrative notebooks are provided at \url{https://github.com/bussilab/MDRefine}.

\section*{Materials and methods}
\texttt{MDRefine} is implemented as a Python package. It requires Python $\geq 3.8$, Numpy, JAX and Scipy libraries. Source code is freely available at \url{https://github.com/bussilab/MDRefine.git}. Tagged versions can be installed from PyPI. The GitHub repository contains documentation as well as a set of illustrative Jupyter notebooks. \texttt{MDRefine} is modular and treats data sets, including both experimental values and quantities from MD simulations, as objects. In the following, we describe the overall architecture of the package and some examples to enable users to make good use of the library.

\subsection*{Structure of \texttt{data} objects}
A complete data set is represented as an object \texttt{data}, which, once defined, is devised to be immutable. Its exact structure is discussed in the Python documentation. The user can directly provide the \texttt{data} object or use the provided
\texttt{load\_data} function to read files saved using the standard Numpy format from a suitably organized hierarchy of directories.
The \texttt{data} object has two main attributes: \texttt{data.properties}, which is a dictionary containing global properties of the data (such as the names of the systems \texttt{sys\_names} and the original forward-model coefficients), and \texttt{data.mol}, which is a dictionary with one item for each molecular system.

For each system, the corresponding item \texttt{data.mol[sys\_name]} (an instance of \texttt{data\_class}) 
contains as attributes both the \textit{experimental values} \texttt{gexp} (measured values $g_{i,exp}$ with estimated uncertainties $\sigma_{i,exp}$) and some quantities evaluated for each frame $x$ of the MD trajectory. The latter includes: the \texttt{weights} $w(x)$, the \textit{observables} \texttt{g} to be compared with $g_{exp}$, and the quantities \texttt{f} required to compute the \textit{force-field corrections} through the method \texttt{ff\_correction}, whose functional form is specified by the user. We include also the possibility to compute some (or all) of the observables \texttt{g} through a user-defined function stored in the method \texttt{forward\_model}, which acts on the attribute \texttt{forward\_qs} (namely, quantities required for the forward model) and the corresponding coefficients. For the sake of clarity, both \texttt{g} and \texttt{gexp} are dictionaries of Numpy arrays split per experiment type \textit{exp\_type}, like for example \texttt{backbone1\_gamma\_3J}, \texttt{backbone2\_beta\_epsilon\_3J}, \texttt{sugar\_3J} (3J scalar couplings computed for several dihedral angles) and \texttt{NOEs} (signals detected through nuclear Overhauser effect), with the attribute \texttt{names} naming each experimental measure. Also \texttt{forward\_qs} is a dictionary of different arrays, but its keys do not necessarily correspond to those of \texttt{g} and \texttt{gexp}. Both \texttt{weights}, \texttt{g[exp\_type]}, \texttt{f} and \texttt{forward\_qs[exp\_type]} are Numpy arrays whose rows correspond to (consecutive) frames $x$ of the MD trajectory.

Vast freedom is left to the \textit{units of measure}, provided the experimental values are loaded consistently with the observables (both \texttt{g} and output quantities of the forward model). Regarding the force-field corrections, the set of \texttt{data.mol[sys\_name]} items include a variable named \texttt{temperature} corresponding to the thermal energy $k_B T$ at which the MD simulation was performed. It is $k_B T = 1$ by default, namely, energies are expressed in unit of temperature, and in this way force-field corrections must be defined. Alternatively, one can express \texttt{temperature} in the preferred unit, as long as the force-field corrections are consistently defined. If the same system has been simulated at multiple temperatures to match different experiments, it is possible to provide the copies as if they were independent systems, each with its own temperature, as long as they are defined with different \texttt{sys\_names}.

Experimental values can be passed either as a ``central reference'' $g_i = g_{i,exp}\pm\sigma_{i}$ or as \textit{upper/lower bound}. The latter is often the
case for data such as \textit{unobserved NOEs}, where the experiment only
provides an upper threshold for the NOE signal below which the signal cannot be revealed $g_i < g_{i,exp}$ with tolerance $\sigma_i$. This is specified by the attribute \texttt{ref} (``reference''), which contains the sign of the inequality to be satisfied ($=$ for central references, $</>$ for upper/lower thresholds respectively), the same sign for each \texttt{data.mol[name\_sys].g[exp\_type]} set of observables. To simplify the implementation, we assume that a tolerance $\sigma$ is always provided and different from zero, altough it can be set to a numerically very small value to reach the regime where the tolerance is neglibible.

We included also the case of experimental values given by free energy differences $\Delta G_{exp}$ between couples of molecules.
We built our implementation based on the prototypical case of computing mutation free energies in single- (S) and double- (D) stranded nucleic acids, which can be measured through denaturation experiments
\cite{piomponi2022molecular}. This has been used in refining force fields for RNA molecules with methylated nucleotides
\cite{piomponi2022molecular, piomponi2024molecular}. In order to compare these experimental measures with MD simulations, alchemical transformations \cite{shirts2007alchemical,piomponi2022molecular} are performed transforming oligomers with wild-type adenosine (A) to methylated ones (M) both in the single strand S and duplex D cases. Free-energy differences resulting from MD simulations can then be compared with experimental values in virtue of the thermodynamic cycle $
\Delta G_{exp, A}-\Delta G_{exp, M} = \Delta G_S - \Delta G_D$. To this aim, we introduced a new class \texttt{data\_cycle\_class} for the thermodynamic cycles, whose instances \texttt{data.cycle[cycle\_name]} contain both the experimental value of $\Delta\Delta G$ (with associated uncertainty) and the temperature. In this case, the \texttt{data.mol} items of the four molecular systems belonging to the cycle \texttt{cycle\_name} have to be indicated in \texttt{infos[`global'][`cycle\_names']} in the same order as \texttt{[`AS', `AD', `MS', `MD']}, where one has the experimental $\Delta G$ between single- (S) and double- (D) -stranded molecules and the alchemical transformation between A and M molecules. In other words, the four systems should be listed in order such that the alchemical transformations are between the first and the third system and between the second and the fourth system,
and the reference experimental value reports the difference between these two alchemical free energies. Each of them has a further attribute \texttt{logZ}, which is the natural logarithm of the partition function shifted by a given reference value and related to the alchemical transformation from A to M molecules (either single- or double-stranded). Fixing a reference ensemble $P_0$ for the energy, then it is $\log Z_A = \log \langle e^{-\beta V_A(x)}\rangle_{P_0}$ and analogous for $\log Z_M$, so the free-energy difference is $\Delta G|_A^M = -k_B T (\log Z_M -\log Z_A)$. In most of the cases, one fixes $P_A$ as reference ensemble $P_0 = P_A$, so $\log Z_A = 0$ and $\Delta G|_A^M = -k_B T \log Z_M = -k_B T \log \langle e^{-\beta\Delta V|_A^M}\rangle_A$, also known as the Zwanzig equation \cite{zwanzig1954high}.

Two specific methods can be used to inform the algorithm about the preference for adjusting the force-field parameters (\texttt{ff\_correction}) and the forward models (\texttt{forward\_model}).
The function \texttt{ff\_correction} takes as input both the full set of force-field correction coefficients $\phi$ specified by \texttt{data.properties.names\_ff\_pars} and the \texttt{f} array previously defined and returns the force-field correction for each frame $\Delta V(x;\phi)$. The function \texttt{forward\_model} takes as input both the full set of forward-model coefficients $\theta$ listed as in \texttt{forward\_coeffs\_0} and the dictionary \texttt{forward\_qs} and returns calculated observables $g_i(x;\theta)$; the output is structured as a dictionary which is going to update \texttt{data.mol[sys\_name].g} that can be compared element-wise against the experimental measures \texttt{data.mol[sys\_name].gexp}.

\subsection*{Basic calculations and reweighting}
Once we have loaded all the required data in the dictionary \texttt{data}, we are now ready to perform basic calculations and reweighting of the ensembles. First of all, we are interested in quantifying how well the ensemble $P_0$ predicted by MD simulations agrees with the experimental measures $g_{i,exp}\pm\sigma_i$. This is usually done using the $\chi^2$, which depends both on the ensemble $P_0$ and on the forward model parameters $\theta$:
\begin{equation}
\chi^2[P_0,\theta]=\sum_i\Bigl(\frac{\langle g_i(x;\theta)\rangle_{P_0}-g_{i,exp}}{\sigma_{i,exp}}\Bigr)^2,
\end{equation}
where $\langle g_i(x;\theta)\rangle_{P_0}$ is the average value of the observable $g_i(x;\theta)$ on the ensemble $P_0$. This is computed by the function \texttt{compute\_chi2}, whose inputs are the \texttt{weights} representative of $P_0$, the observables \texttt{g} $g_i(x;\theta)$, the experimental values \texttt{gexp} and the \texttt{ref} dictionary previously defined. This last variable allows for including also the case of experimental values for upper/lower bounds. For example, the values of the item \texttt{gexp[`uNOEs']} (unobserved NOEs) are upper bounds $g_{i,exp}$ for the signal observables \texttt{g[`uNOEs']} with a tolerance $\sigma_i$ (\texttt{ref[`uNOEs'] = `$<$'}) and the corresponding contribution to the $\chi^2$ is

\begin{equation}
\sum_{i\in uNOEs}\Bigl[\max\Bigl(0,\frac{\langle g_i(x)\rangle_{P_0}-g_{i,exp}}{\sigma_{i,exp}}\Bigr)\Bigr]^2.
\end{equation}
The output values of the function \texttt{compute\_chi2} are: the average quantities $\langle g_i(\theta;x)\rangle_{P_0}$, the $\chi^2$ for each observable $i$, the relative difference $\frac{\langle g_i(x)\rangle_{P_0}-g_{i,exp}}{\sigma_{i,exp}}$ and the total $\chi^2$.

If there is a significant discrepancy, the reference ensemble $P_0$ might require some adjustments, either by varying the empirical coefficients $\phi$ of the force field or by refining the ensemble $P_0$ with full flexibility, or both \cite{gilardoni2024boosting}. Once the dimensionless \textit{correction} $\beta\Delta V(x)$ (namely, the logarithm of the correction factor) for each frame is given, we can reweight the ensemble by employing the function \texttt{compute\_new\_weights}, which takes as input the original weights $w_0(x)$ and the correction $\beta\Delta V(x)$ and returns the new (normalized) weights
\begin{equation}
w(x)=\frac{1}{Z} w_0(x) \, e^{-\beta \Delta V(x)}
\label{eqn:upp_low}
\end{equation}
together with $\log Z$. For example, the correction might be the output of \texttt{ff\_correction} with some coefficients $\phi$ reasonably chosen.

Once we have got the new (reweighted) ensemble $P$, represented by the weights $w(x)$, we can both compute the new $\chi^2[P]$ and compare $P$ with the original ensemble $P_0$. In the case of alchemical calculations, the function \texttt{compute\_DeltaDeltaG\_terms} allows for computing the new free-energy differences $\Delta G|_{A\phi}^{M\phi}$ and $\chi^2$ for the $\Delta\Delta G$ of the thermodynamic cycle by passing as input variables the \texttt{data} object and the dictionary of $\log Z_\phi = \log \langle e^{-\beta\Delta V(x;\phi)}\rangle_\phi$ values for the reweighted ensembles $P_\phi$ of each molecular system in the cycle.

Regarding the comparison of $P$ with the reference ensemble $P_0$,
if $P$ is determined by reweighting $P_0$ with some force-field corrections $\Delta V(x;\phi)$, we can evaluate the deviation just by computing the $L^2$ distance in the parameter space $\sum_i \phi_i^2$ (\texttt{l2\_regularization}). However, changing by the same quantity some coefficients $\phi_j$ rather than others may lead to different variations of the ensemble, and the $L^2$ norm itself is undefined if the
coefficients are expressed in different units.
An alternative possibility without these problems is the 
Kullback-Leibler divergence:
\begin{equation}
D_{KL}[P|P_0] = \sum_x P(x) \log \frac{P(x)}{P_0(x)}
= -\beta\langle\Delta V(x) \rangle_P - \log Z
\end{equation}
with $ Z = \langle e^{-\beta\Delta V(x)} \rangle_0$ the partition function of $P$, whose logarithm is returned by \texttt{compute\_new\_weights}. 
This quantity is the negative of the relative entropy between the ensemble $P$ and $P_0$ and is positively definite, the greater it is the further the ensemble $P$ is from $P_0$. It can be computed through the function \texttt{compute\_D\_KL}, whose inputs are the new weights $P(x)$, the correction $\Delta V(x)$, the temperature $\beta^{-1}$ and $\log Z$. Now we have all the basic ingredients to compute the loss functions defined in the next section.

\subsection*{Loss functions and search for optimal solutions}

The simplest option for refining an ensemble is to search for the ensemble which best fulfils the trade-off between the the experimental measures and the initial hypothesis $P_0$. Building on the maximum entropy principle, this \emph{Ensemble Refinement} procedure can be performed by minimizing the following loss function \cite{hummer2015bayesian}:

\begin{equation}
\mathcal L[P]=\frac{1}{2}\chi^2[P]+\alpha D_{KL}[P|P_0],
\end{equation}
where $\alpha>0$ is a hyperparameter regulating the trade-off. One can show the optimal ensemble $P(x)$ takes the functional form
\begin{equation}
P_*(x)=\frac{1}{Z_\lambda} P_0(x) e^{-\vec\lambda\cdot\vec g(x)}
\end{equation}
where
\begin{equation}
\lambda_i = \frac{\langle g_i\rangle_{P_*}-g_{i,exp}}{\alpha\sigma_{i,exp}^2}.
\end{equation}

Hence, we can exploit this dimensionality reduction and search for the optimal solution in the subspace described by the parameters $\lambda$ (one for each experimental value). Minimizing $\mathcal L[P_\lambda]$ turns out to be equivalent (i.e., same extreme point and corresponding value of the function) to maximize $-\alpha\Gamma(\lambda)$ \cite{frohlking2023simultaneous,kofinger2019efficient}, being
\begin{equation}
\Gamma(\lambda) = \log Z_\lambda + \vec\lambda\cdot\vec{g}_{exp} + \frac{1}{2} \alpha \sum_i \lambda_i^2 \sigma_i^2.
\end{equation}

Therefore, we reduce to determining the optimal ensemble $P_*$ by minimizing $\Gamma(\lambda)$.
Such \emph{Ensemble Refinement}
is implemented in the \texttt{minimizer} function.
Notice that a similar functionality is also available in other packages, \emph{e.g.},
\url{https://github.com/KULL-Centre/BME}\cite{bottaro2020integrating} and \url{https://github.com/bio-phys/BioEn}\cite{hummer2015bayesian}).
A prototype implementation is also available at \url{https://github.com/bussilab/py-bussilab/blob/master/bussilab/maxent.py}. The \texttt{gamma\_function} here implemented takes as input the $\lambda$, \textit{g}, \textit{gexp}, \textit{weights}, all as Numpy arrays, and the hyperparameter $\alpha$, and returns the value of $\Gamma(\lambda)$.
Notice that this value coincides with $\mathcal L[P]$ at the optimal point $P_*$.
The $\Gamma$ function can be shown to be convex in the parameters $\lambda$, at variance with the function 
$\mathcal L[P(\lambda)]$. This allows for a more robust numerical optimization.

The dimensionality reduction with $\lambda$ parameters holds whenever the loss function has the form $\mathcal L[P] = \ell (\langle g_i(x)\rangle_P) + \alpha D_{KL}[P|P_0]$ with $\ell(\langle g_i(x)\rangle_P$ a generic function of the average values. Still, the derivation of the $\Gamma$ function, enabling a more efficient  minimization, requires $\ell$ to be a quadratic function of its arguments. When minimizing $\Gamma$, experimental values for upper/lower bounds (as in Eq. \ref{eqn:upp_low}) can be included by employing boundaries in the $\lambda$ space ($\lambda_i\geq 0$ for upper limits \texttt{ref = `$<$'}, $\lambda_i\leq$ 0 for lower limits \texttt{ref = `$>$'}).

The loss function described above is the one used in standard ensemble refinement procedures.
When comparing MD simulations with experiments, all the errors, both coming from the ensembles, the force field and the forward model, come into play. Hence, we consider the possibility to simultaneously refine all of them. 
To this aim, and motivated by previous works in these directions
\cite{frohlking2023simultaneous,gilardoni2024boosting}, we deem the following loss function to be suitable for our task:

\begin{equation}
\mathcal{L}(P,\phi,\theta)=\frac{1}{2}\chi^2[P,\theta]+\alpha D_{KL}[P|P_\phi]+\beta R_1[P_\phi|P_0]+\gamma R_2[\theta|\theta_0]
\label{eqn:loss_complete}
\end{equation}
where $\alpha,\beta,\gamma$ are hyperparameters regulating the strength of the corresponding regularization terms, $P_\phi$ is the canonical ensemble determined by applying the force-field correction $\phi$ to the reference ensemble $P_0$, and $R_1,\,R_2$ are customizable regularizations for the force-field and the forward-model corrections, respectively. The loss function is considered to be implicitly summed over all the investigated molecular systems. This task can be interpreted as a Pareto (or multi-objective) optimization \cite{kofinger2024encoding}, solved through a linear scalarization, with $\alpha,\beta,\gamma \geq0$ the relative weights of the objective functions. This loss function is quite adaptable to different aims by formally setting some of the hyperparameters to $+\infty$. For example, if only $\beta$ is finite we recover force-field refinement.
For $\alpha<+\infty$, $\beta<+\infty$, and $\gamma=+\infty$ we fully trust the original forward-model coefficients getting back to \cite{gilardoni2024boosting}.
Similarly, for $\alpha<+\infty$, $\beta=+\infty$, and $\gamma<+\infty$ we recover \cite{frohlking2023simultaneous}.

To minimize this loss function, we exploit the same dimensionality reduction previously introduced (based on $\lambda$ coefficients), except that now it appears in a nested optimization. In other words, we determine the optimal $\lambda^*(\phi,\theta)$ (by minimizing $\Gamma(\lambda;\phi,\theta)$ at fixed $\phi,\theta$ -- inner minimization) at any step of the (outer) minimization in the $\phi,\theta$ parameters space. 
Making the dependence $\lambda^*(\phi,\theta)$ explicit in the loss function $\mathcal L(P,\phi,\theta)$ we get
\begin{equation}
\mathcal{L}_1(\phi,\theta) = \mathcal{L}(\lambda^*(\phi,\theta),\phi,\theta)=-\alpha\Gamma(\lambda^*(\phi,\theta);\phi,\theta)+\beta R_1(\phi)+\gamma R_2(\theta).
\end{equation}
and the optimal coefficients are determined by
\begin{equation}
(\phi^*,\theta^*) = \arg\min_{\phi,\theta} \mathcal L_1(\phi,\theta) \,\,\,\,\, \mathrm{and} \,\,\,\,\, \lambda^* = \arg\min_\lambda \Gamma(\lambda;\phi^*,\theta^*)
\end{equation}
(see \textit{Supplementary - Computation of derivatives} and \textit{Implementation of \texttt{minimizer}} for further details).

Besides the \texttt{data} object, we have to define also the regularization terms. These two terms are encoded in the \texttt{regularization} dictionary, with two items: \texttt{force\_field\_reg}, which can be either a string (\texttt{KL divergence}, \texttt{plain l2}) or a user-defined function acting on $\phi$; \texttt{forward\_model\_reg}, which is a user-defined function acting on $\theta$ and $\theta_0$ (for example, an $L^2$ regularization). Then, the \texttt{loss\_function} is minimized following the described strategy by the \texttt{minimizer} routine. In this way, passing to \texttt{minimizer} the \texttt{data} object, the \texttt{regularization} dictionary and the hyperparameters, we get an output object \texttt{Result} containing the optimal solution $(P, \phi, \theta)$ and its properties.
By default, the hyperparameters $\alpha,\beta,\gamma$ are set to $+\infty$, implying no refinement in the corresponding directions.

\subsection*{Searching for optimal hyperparameters via cross validation}

Suitable ranges of the hyperparameters $\alpha,\beta,\gamma$ can be determined by searching for an intermediate range between underfitting and overfitting. This can be done by splitting the data set into training and validation set and then searching for the hyperparameters which minimize the $\chi^2$ on the validation set (cross validation). This splitting can be done in several ways.
The way we recommend is to select some frames and some observables to belong to the validation set, leaving them out from the search of the optimal coefficients, which is done using only the training set. Keeping frames in the validation set makes sure the corrections can be generalized to a new (or longer) trajectory simulated for the same system, whereas keeping observables in the validation set strengthen the transferability of the correction to new observables \cite{frohlking2022automatic}.

The function \texttt{select\_traintest} performs the splitting following these principles. Passing the \texttt{data} object, it splits the data set into training and validation set, with default ratios of the validation frames \texttt{test\_frames\_size} and validation observables \texttt{test\_obs\_size} of 20\% on the full number of frames and observables for each molecular system. Alternatively, one can also pass two dictionaries \texttt{test\_frames} and \texttt{test\_obs} with user-defined values. The input variable \texttt{random\_state} (an integer number) allows for fixing the seed used in the random splitting, thus making it reproducible. By default, the validation observables are defined only on validation frames, however they can be defined on all the frames just by setting \texttt{if\_all\_frames = True}. %

The \texttt{validation} function, called by \texttt{minimizer}, allows for including in the output object \texttt{Result} also properties referring to the validation set, such as the $\chi^2$, both for training observables on validation frames and for validation observables. %

With the implementation described so far,
one can scan over different values of the hyperparameters, minimize the loss function on the training set and evaluate the $\chi^2$ on the validation set,
to identify values of the hyperparameters that maximize transferability of the parameters.
Whereas this procedure allows for determining suitable values for the hyperparameters, it suffers from the curse of dimensionality: while an exhaustive search is feasible for 1 or 2 dimensions, making a scan in 3 dimensions or more is cumbersome. To overcome this issue, we implemented an automatic search of the optimal hyperparameters\cite{lorraine2020optimizing}.

This search is performed by randomly splitting the full data set into training and validation set several times (as indicated by \texttt{random\_states}) and minimizing the average $\chi^2$ on the validation set, evaluated at the optimal solution found in training. Since both $\phi$ and $\theta$ appear together in the minimization, in the following we will use $\mu=(\phi,\theta)$ as a shortcut. The optimal coefficients depend on the hyperparameters via
\begin{equation}
\begin{split}
&\lambda^*(\mu^*;\alpha)=\arg\min_\lambda \Gamma(\lambda;\mu^*,\alpha)\\
&\mu^*(\alpha,\beta,\gamma)=\arg\min_\mu \mathcal L_1(\mu;\alpha,\beta,\gamma),
\end{split}
\end{equation}
therefore by the chain rule the derivatives of the $\chi^2(\lambda^*,\mu^*)$ with respect to the hyperparameters result
\begin{equation}
\frac{\partial\chi^2}{\partial\alpha}
= \frac{\partial\chi^2}{\partial\mu_i}\frac{\partial\mu_i^*}{\partial\alpha} + \frac{\partial\chi^2}{\partial\lambda_i}\Bigl(\frac{\partial\lambda_i^*}{\partial\mu_j}\frac{\partial\mu_j^*}{\partial\alpha}+\frac{\partial\lambda_i^*}{\partial\alpha}\Bigr)
\end{equation}

\begin{equation}
\frac{\partial\chi^2}{\partial(\beta,\gamma)}
= \frac{\partial\chi^2}{\partial\mu_i}\frac{\partial\mu_i^*}{\partial(\beta,\gamma)} + \frac{\partial\chi^2}{\partial\lambda_i}\frac{\partial\lambda_i^*}{\partial\mu_j}\frac{\partial\mu_j^*}{\partial(\beta,\gamma)}
\end{equation}
(see \textit{Supplementary - Computation of derivatives} for the complete derivation).

This optimization is performed by \texttt{hyper\_minimizer}, which requires the \texttt{data} object, the starting values of the hyperparameters which, by default are constrained to be $+\infty$, the \texttt{regularization} and the \texttt{random\_states}. One can also choose among minimizing the $\chi^2$ for training observables and validation frames (\texttt{which\_set = `validation'}) or validation observables (\texttt{which\_set = `test'}). Since the minimization acts on orders of magnitude (namely, in logarithmic scale), the optimal hyperparameters are returned as $\log_{10}(\alpha,\beta,\gamma)$. 

\subsection*{All at once: \texttt{MDRefinement}}

\begin{figure}
    \includegraphics[width=\linewidth]{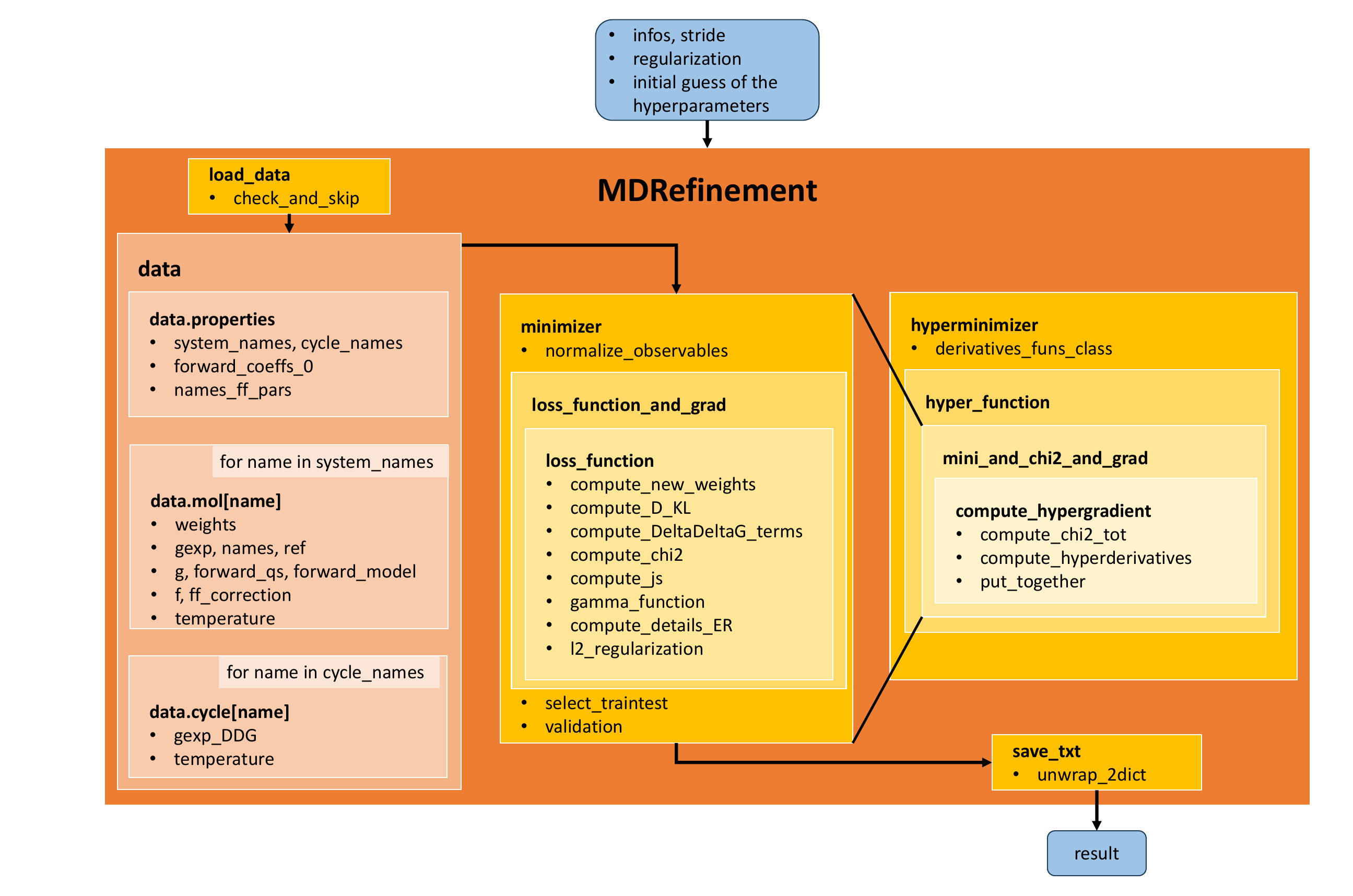}
    \caption{Software architecture: \texttt{MDRefinement} contains two main blocks \texttt{minimizer} and \texttt{hyperminimizer}, besides \texttt{load\_data} and \texttt{save\_txt}. The input variables of \texttt{MDRefinement} include \texttt{infos} (dictionary with information about the data set to be loaded), user-defined \texttt{regularization} terms and initial guess for the hyperparameters (selected refinement).}
    \label{fig:software_architecture}
\end{figure}

\begin{figure}
    \centering
    \includegraphics[width=1\linewidth]{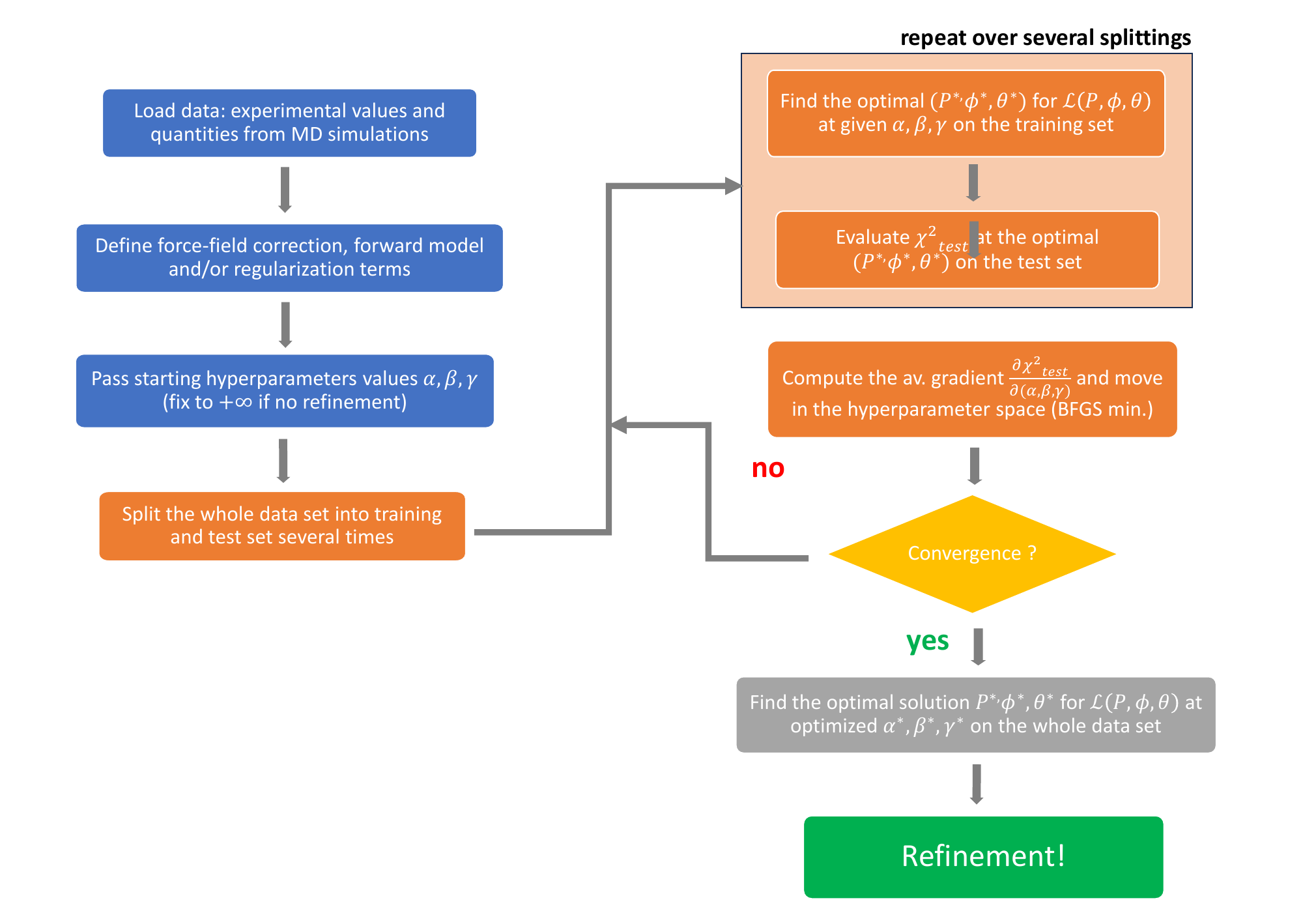}
    \caption{Flow chart of \texttt{MDRefinement}: after loading the data and defining the functions (blue blocks), the cross-validation section (in orange) determines the optimal hyperparameters, which are then used (gray block) to get the most suitable refinement on the data set.}
    \label{fig:flow_chart}
\end{figure}

A single function is defined, named \texttt{MDRefinement}, which encompasses all the functions defined so far, as illustrated in Fig. \ref{fig:software_architecture}. Its flow chart is shown in Fig. \ref{fig:flow_chart}. It loads both the \texttt{data} (from input files according to \texttt{infos}) and the \texttt{regularization}, search for the optimal hyperparameters in cross validation and returns the optimal solution \texttt{Result} on the full data set. The output values are also saved as \texttt{csv} files in a folder named \texttt{results}, including \texttt{input}, \texttt{min\_lambdas}, \texttt{result} object, the steps in the hyperparameter search \texttt{hyper\_search} and Numpy arrays with the optimized weights for each of the simulated frames.

\section*{Results}

To showcase the capabilities of the \texttt{MDRefine}, we present the analysis of a number of precomputed trajectories. Part of the data replicates results presented
in previous paper \cite{piomponi2022molecular,frohlking2023simultaneous,gilardoni2024boosting}.
New analysis are also presented that take advantage of the new capabilities
introduced in this package.
Full Python notebooks that can be used to reproduce these analyses are
available at \url{https://github.com/bussilab/MDRefine}.

\subsection*{Force-field refinement with denaturation experiments}
In this section we reproduce force-field corrections derived in \cite{piomponi2022molecular} and discuss the effect of using a different regularization strategy based on the relative entropy. This is performed in the notebook \texttt{Tutorial.ipynb}, Section 2b. The main intent of this refinement is to improve the description of N6-methyladenosine (m$^6$a), one of the most common post-transcriptional modifications of RNA, consisting in the methylation of the standard adenosine at position N6. Notably, m$^6$a can exist in two possible conformations, depending on the orientation of the methyl group with respect to the rest of the nucleobase: the \emph{syn} and \emph{anti} isomers, which are distinguished by the value of the torsional angle $\eta_6$ defined by the atoms N1-C6-N6-C10. The \emph{syn} conformation is known to be the most stable for unpaired m$^6$a, whereas the \emph{anti} isomer is the one expected when m$^6$a is Watson-Crick paired with uracil in an internal position of a double-stranded RNA.

As a starting parametrization we used the modrna08 force-field\cite{aduri2007amber}, developed by Aduri and coworkers
using the standard AMBER protocol based on quantum-mechanical calculations and ReSP (Restrained Electrostatic Potential) fitting \cite{bayly1993well}.
This parametrization does not predict the correct \emph{syn}/\emph{anti} balance for unpaired m$^6$a. We then apply a correction to the force-field using the same prescription
outlined in \cite{piomponi2022molecular}. This correction includes a sinusoidal term acting on the $\eta_6$ torsional angle $\Delta V(\eta_6)=-V_\eta \cos\eta_6$ (positive values of $V_\eta$ favour syn conformations over anti) and a reparametrization of the partial charges close to the methyl group (C6, N6, H61, N1, C10, H101/2/3), with the constraint on the total charge to be constant $\sum_i \Delta Q_i =0$ and on the charges of the three equivalent hydrogens (H101/2/3) to be identical. Here $\Delta Q_i$ represents the variation of the partial charge of atom $i$
when compared with the starting parametrization. This results in 6 free parameters: $V_\eta$ and 5 partial charges $\Delta Q_i$. 

The total energy depends quadratically on the charge perturbations $\Delta Q_i$, hence, following \cite{piomponi2022molecular}, it can be expressed as
\begin{equation}
\Delta V(x) = \sum_{i=1}^5 K_i(x) \Delta Q_i + \sum_{i=1}^5 \sum_{j=1}^5 K_{ij}(x) \Delta Q_i \Delta Q_j - V_\eta \cos\eta_6(x).
\end{equation}
The 20 coefficients $K_i$ and $K_{ij}$ can be determined through linear algebra operations starting from 20 linearly independent sets of charge perturbations, allowing the computation of the total energy for any choice of the charges $\Delta Q_i$ without the need to repeat the computationally expensive electrostatic-energy calculation.

These 6 free parameters ($\Delta Q_i, \, V_\eta$) were fitted by comparing the reweighted predictions of MD simulations with a number of denaturation experiments measuring free-energy differences $\Delta G$. The molecules used in the training phase include: m$^6$a (\emph{syn}/\emph{anti} difference), $\nobarfrac{\mathrm{UACG6CUG}}{\mathrm{AUGCUGAC}}$, $\nobarfrac{\mathrm{CGAU6GGU}}{\mathrm{GCUAUCCA}}$, $\nobarfrac{\mathrm{6CGC}}{\mathrm{\,\,\,GCG}}$, $\nobarfrac{\mathrm{GCG6}}{\mathrm{CGC\,\,\,}}$ where 6 stands for the m$^6$a nucleoside. This comparison is made by realizing a thermodynamic cycle through alchemical calculations (see section \textit{Structure of} \texttt{data} \textit{objects} and \cite{shirts2007alchemical,roost2015structure,piomponi2022molecular} for further details). The corresponding loss function includes a generic regularization term $R(\Delta Q, V_\eta)$, which encodes the prior knowledge on the force field:
\begin{equation}
\mathcal L(\Delta Q, V_\eta) = \frac{1}{2}\sum_{i=1}^{N_{cycles}} \Bigl(\frac{\Delta\Delta G_i(\Delta Q, V_\eta)-\Delta\Delta G_{i,exp}}{\sigma_{i,exp}}\Bigr)^2 + \beta R(\Delta Q, V_\eta).
\end{equation}

Strictly speaking, the $L^2$ regularization requires dimensionally homogeneous parameters, which is not the case here.
In \cite{piomponi2022molecular} it has been suggested that, for this system, it suffices to regularize the partial charges.
We follow the same protocol here, resulting in a $L^2$ regularization that reads:
\begin{equation}
R(\Delta Q, V_\eta)=\sum_{i=1}^6 \Delta Q_i^2 = \sum_{i=1}^5 \Delta Q_i^2 + \Bigl(\sum_{i=1}^5 \Delta Q_i + 2\Delta Q_{H101}\Bigr)^2.
\end{equation}

\begin{figure}
    \captionsetup{justification=justified, singlelinecheck=off}
    \begin{subfigure}[b]{0.4\textwidth} %
        \caption[]{$L^2$ regularization, $\beta=10 e^{-2}$}%
        \includegraphics[width=\textwidth]{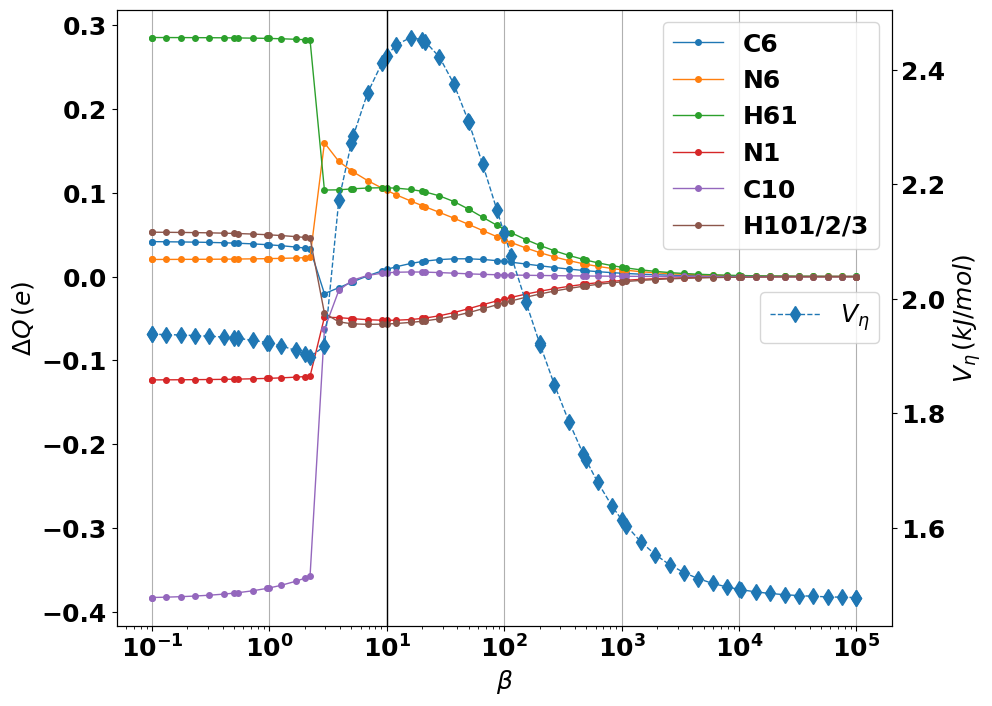}

    \end{subfigure}
    \begin{subfigure}[b]{0.4\textwidth}  %
        \caption[]{$D_{KL}$, $\beta=0.05$}%
        \includegraphics[width=\textwidth]{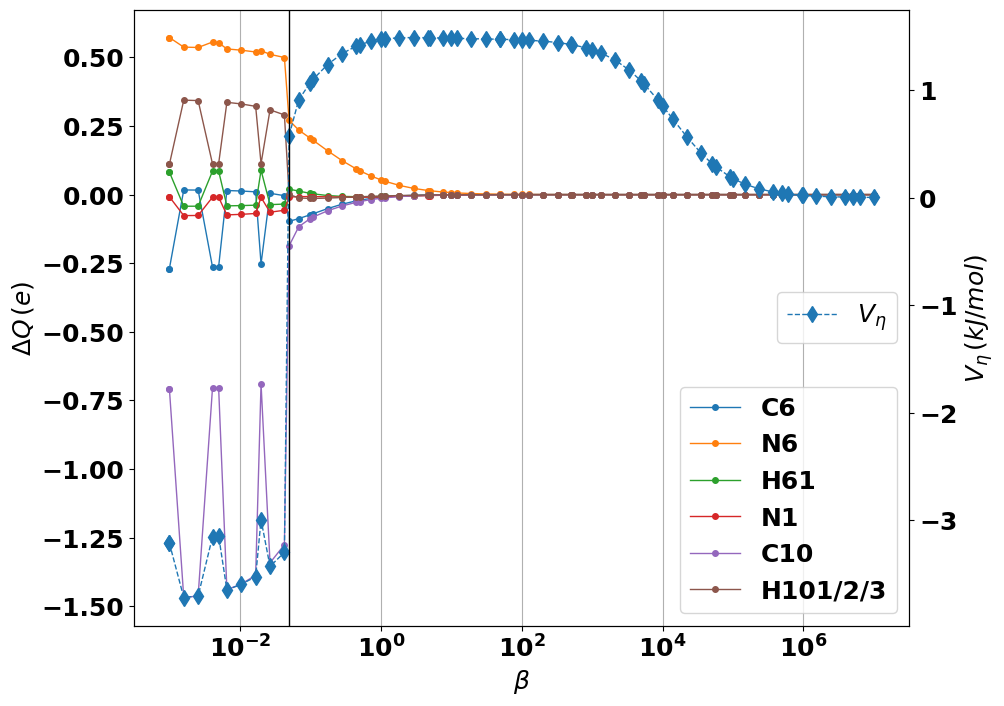}

    \end{subfigure}
    \vskip\baselineskip
    \begin{subfigure}[b]{0.4\textwidth}   %
        \caption[]{relative Kish size} %
        \includegraphics[width=\textwidth]{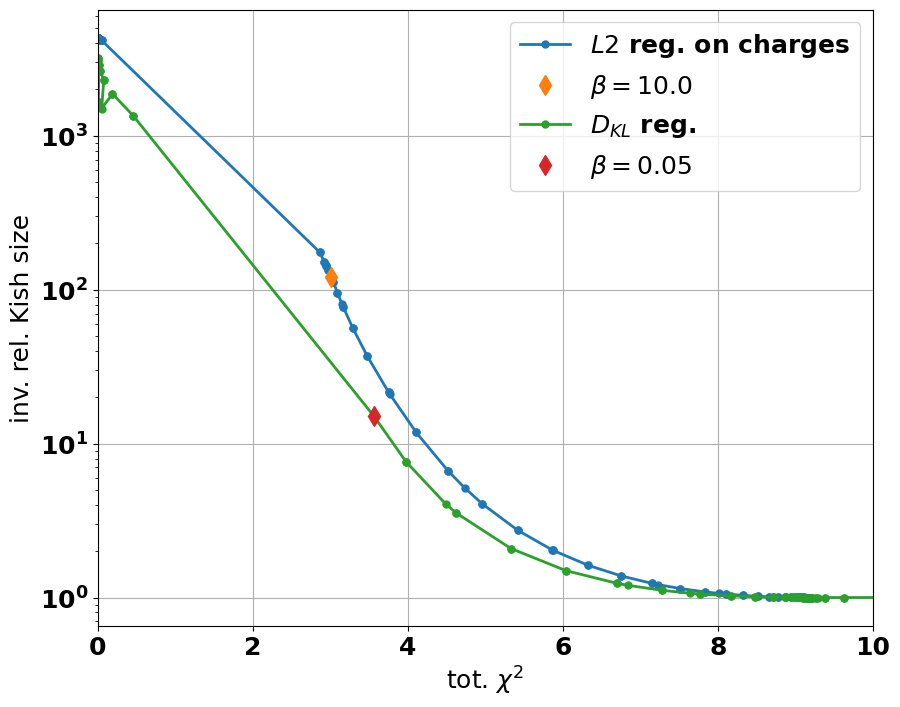}

    \end{subfigure}
    \begin{subfigure}[b]{0.4\textwidth}   %
        \caption[]{partial charges}%
        \includegraphics[width=\textwidth]{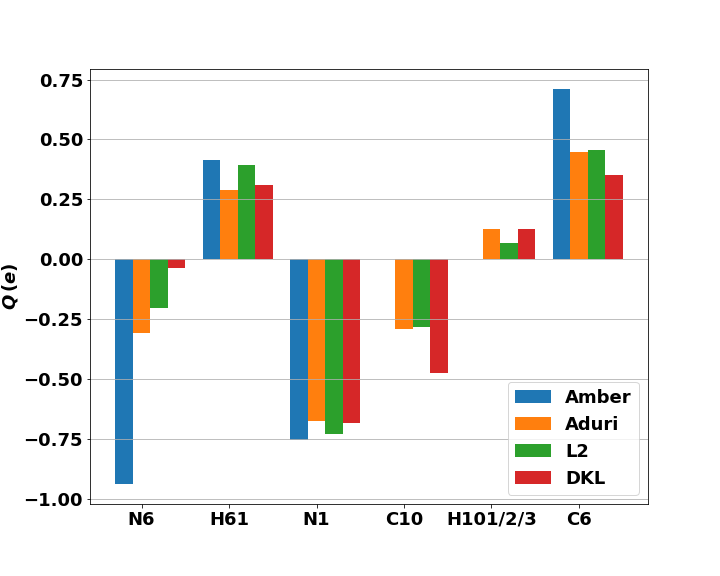}
        
    \end{subfigure}
    \caption[ alchemical ]
    {\small Refinement of charges and $V_\eta$ for alchemical calculations; comparison between $L^2$ (a) and relative entropy (b) regularization. (c) Inverse of the relative Kish size $K^{-1}=\sum_x w(x)^2/w_0(x)$, measuring the deviation from the reference ensemble, as a function of the total $\chi^2$ of the $\Delta\Delta G$ on the thermodynamic cycles, at constant steps in the $\log\beta$ scale.
    (d) Partial charges $Q(e)$ for different parametrizations: Amber force field for wild-type adenosine, Aduri force field for methylated adenosine, $L^2$ and relative entropy regularization.}
    \label{fig:alchemical_refinement}
\end{figure}

As an alternative procedure, we here test the possibility of regularizing
using the relative entropy $R(\Delta Q, V_\eta)=D_{KL}[P(\Delta Q, V_\eta)|P_0]$
\cite{kofinger2021empirical},
where the sum over all the systems is implicit.
The results are shown in Fig. \ref{fig:alchemical_refinement}.

In both cases (Fig. \ref{fig:alchemical_refinement} a, b), starting from a high value of the hyperparameter $\beta$ and decreasing it, the parameters change abruptly below a certain threshold. This can be interpreted as a transition to overfitting the experimental data.
Fig. \ref{fig:alchemical_refinement}c reports the
relationship between the discrepancy from the reference ensemble
and the discrepancy from the experiment, using both regularization flavors.
In both cases, a smaller discrepancy from experiment ($\chi^2$) can
only be obtained at the price of a larger perturbation of the ensemble.
As expected, relative entropy regularization results in less perturbed ensembles
than $L^2$ regularization for the same $\chi^2$ value.
Interestingly, even if the $\beta$ value are uniformly spaced in logarithmic scale,
an abrupt change to the overfitting regime is observed, leading to a sharp decrease in $\chi^2$.

Hence, we choose hyperparameter values slightly before this threshold: $\beta=10\,e^{-2}$ (the same as in Ref.~\cite{piomponi2022molecular} ) and $\beta\simeq 0.5$ for $L^2$ and $D_{KL}$ regularizations, respectively. These values are marked with bold vertical line in Fig. \ref{fig:alchemical_refinement} a, b. Notice that choosing the optimal hyperparameter on the basis of the plots comparing deviation from theory and deviation from experiments has been suggested in some recent works \cite{kofinger2024encoding}.
Having an objective criterion to choose the hyperparameter is critical
in this case because it allows us to compare the results obtained with
the two regularization strategies, even if the value of the hyperparameters
are not directly comparable. An alternative strategy for
choosing hyperparameters that we didn't test here is cross validation,
as discussed below.

In Fig.~\ref{fig:alchemical_refinement} we report the partial charges
obtained for the chosen hyperparameters, and compare them with other sets
of partial charges for the same system.
There are important differences when comparing the refinement with $L^2$ or relative entropy regularization. Most importantly, the $D_{KL}$ regularization leaves relatively unconstrained C6 and N6 charges. This is due to the fact that these atoms can not form close contacts with other atoms, and thus their charge can be modified without a significant perturbation of the ensemble. The strengthening of Watson-Crick hydrogen bonds that when using $L^2$ regularization was obtained by making N1 and H61 slightly more polar, is here obtained by making N6 virtually apolar, transferring its charge largely to C10. Indeed, the $L^2$ regularization keeps the final parameters close to those chosen based on the quantum mechanical calculations in Ref.~\cite{aduri2007amber}, whereas $D_{KL}$ considers the effect of the parameters on the resulting ensemble.

\subsection*{Simultaneous forward-model and ensemble refinement}

In this section we reproduce results from \cite{frohlking2023simultaneous} (see \texttt{Tutorial\_3.ipynb}). The main purpose is to better estimate the empirical coefficients of the Karplus equations, relating the dihedral angles to the $^3J$ couplings measured in nuclear magnetic resonance experiments. This refinement is made by taking into account at the same time possible inaccuracies in the structural ensembles. By looking at our loss function (Eq. \ref{eqn:loss_complete}), this choice corresponds to setting $\beta=+\infty$, i.e., combining the refinement of the forward model with that of the ensembles, 
without further flexibility in the force-field parameters.

In our case study, the molecular systems include several RNA oligomers: AAAA, CAAU, CCCC, GACC, UUUU, UCAAUC. For each of them, we employ some experimental measures such as the $^3J$ scalar couplings for some dihedral angles, the NOE signals and the upper bound values of unobserved NOE signals. MD simulations have been performed \cite{frohlking2023simultaneous} with the standard AMBER BSC0 OL3 RNA force-field\cite{cornell1996second, wang2000well, perez2007refinement, zgarbova2011refinement} with the van der Waals modification of phosphate oxygens \cite{steinbrecher2012revised}; the water model is OPC\cite{izadi2014building}. The sampling has been enhanced with parallel tempering
(see Ref.~\cite{frohlking2023simultaneous} for further details).

The forward models here consist in the Karplus equations with empirical parameters $A, B, C$, relating the dihedral angle $\phi$ to the $^3 J$ scalar coupling:
\begin{equation}
^3J(\phi) = A\cos^2 \phi + B\cos\phi + C.
\end{equation}
The employed dihedral angles are: $\beta$ (P-O5'-C5'-H5', P-O5'-C5'-H5''), $\varepsilon$ (H3'-C3'-O3'-P), $\gamma$ (H5'-C5'-C4'-H4', H5''-C5'-C4'-H4') (all along the backbone) and the sugar puckering angles $\nu_1, \nu_2, \nu_3$ (H1'-C1'-C2'-H2', H2'-C2'-C3'-H3', H3'-C3'-C4'-H4').  
We use as reference coefficients the values reported in \cite{lankhorst1985carbon, davies1978conformations, condon2015stacking}. The same coefficients were used for $\beta$ and $\varepsilon$ dihedral angles, both corresponding to H-C-O-P atom types. The regularization of the forward model here adopted is the square deviation between new and reference $^3J$ couplings averaged over the angle, which can be obtained from the parameters as:
\begin{equation}
\begin{split}
& R(A,B,C) = \frac{1}{2\pi}\int_0^{2\pi} d\phi\,(^3J(\phi)-^3 J_0(\phi))^2 \\
& = \frac{3}{8}(A-A_0)^2 + \frac{1}{2}(B-B_0)^2 + (C-C_0)^2 + (A-A_0)(C-C_0),
\end{split}
\end{equation}
This expression is identical to the one used in 
Ref.~\cite{frohlking2023simultaneous}, except for an arbitrary prefactor 2.

\begin{figure}
    \centering
    \includegraphics[width=1\linewidth]{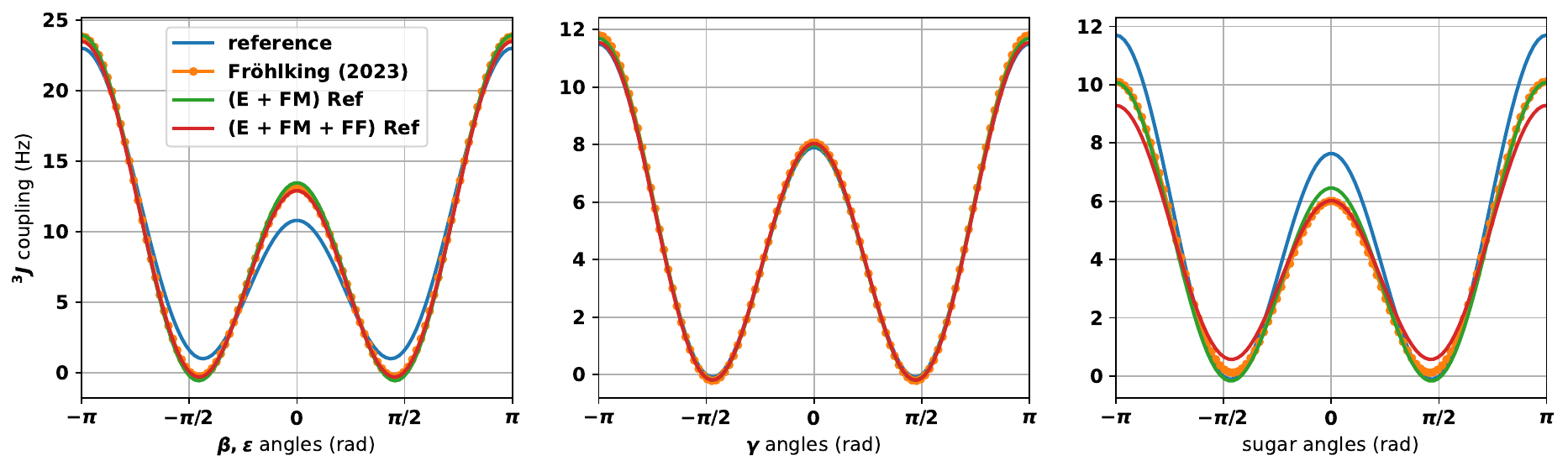}
    \caption{Refinement of Karplus equations. Comparison between: reference values (Lankhorst\cite{lankhorst1985carbon} for $\beta,\epsilon$ angles, Davies\cite{davies1978conformations} for $\gamma$ angles, Condon\cite{condon2015stacking} for sugar angles), refinement by Fröhlking et al.\cite{frohlking2023simultaneous}, this work with Ensemble + Forward-Model Refinement \textit{(E + FM) Ref} and with the fully combined approach \textit{(E + FM + FF) Ref}.}%
    \label{fig:Karplus_refinement}
\end{figure}

We then perform a simultaneous optimization of the parameters $A$, $B$, and $C$ (three for each of the three Karplus equations), and of the structural ensembles, as in \cite{frohlking2023simultaneous}.
To search for optimal hyperparameters, we perform an
 automatic search with the \texttt{hyper\_minimizer} tool described
 above through minimization of the $\chi^2$ on test observables.
  Notice that the cross-validation performed in \cite{frohlking2023simultaneous} employed a slightly different choice for the training/test set, hence we do not expect to get exactly the same hyperparameter values as the optimal ones.
  In particular, the values corresponding to those reported in \cite{frohlking2023simultaneous}
  are $\alpha=174.33,\,\gamma=6.87\,Hz^{-2}$, whereas those obtained in this
  work are $\alpha=46.31,\,\gamma=0.96\,Hz^{-2}$.
  Then, we compare the parameters found for these two sets of optimal hyperparameters with both the reference values and those determined in \cite{frohlking2023simultaneous} (Fig. \ref{fig:Karplus_refinement_compare}  and Table \ref{tab:karplus} in Supplementary). Overall, our results adjust the reference empirical coefficients of the Karplus equations in the same direction suggested in \cite{frohlking2023simultaneous}. Moreover, the refinement performed with hyperparameters optimized by \texttt{MDRefine} is closer to the results proposed in \cite{frohlking2023simultaneous}. Hence, we report the refinement with this second choice of coefficients in Fig. \ref{fig:Karplus_refinement}.

Remarkably, these refinements minimally modify the structural ensembles: what matters most is the inaccuracy in the forward model. Indeed, the Kullback-Leibler divergence between reweighted and original ensembles is quite small, and also the histograms of the angles are not significantly modified (see Supplementary). Hence, we conclude that most of the discrepancy between MD trajectories and experiments
is a consequence of forward-model errors rather than of genuine structural discrepancies.

\subsection*{Fully combined refinement}

In this section we describe an application of the fully combined refinement,
which adds the adjustment of the force-field coefficients to the simultaneous refinement of ensembles and forward models shown in the previous Section. This is obtained by minimizing the full loss function in Eq. \ref{eqn:loss_complete}. In this way, we are able to smoothly interpolate between these three different refinements (force field, ensembles and forward model), casting in a single framework several independent works \cite{frohlking2023simultaneous, piomponi2022molecular, gilardoni2024boosting}.
The benchmark case discussed here is fully described in \texttt{Tutorial\_3.ipynb}, and builds on the refinement described in the previous section adding some flexibility in the force-field coefficients. In particular, following \cite{gilardoni2024boosting}, we focus on a sinusoidal correction on $\alpha$ dihedral angles
\begin{equation}
\Delta V(\alpha)=\phi_1 \cos\alpha + \phi_2\sin\alpha.
\end{equation}

\begin{figure}
    \captionsetup{justification=justified, singlelinecheck=off}
    \begin{subfigure}[b]{0.5\textwidth} %
        \caption[]{force-field correction}%
        \includegraphics[width=\textwidth]{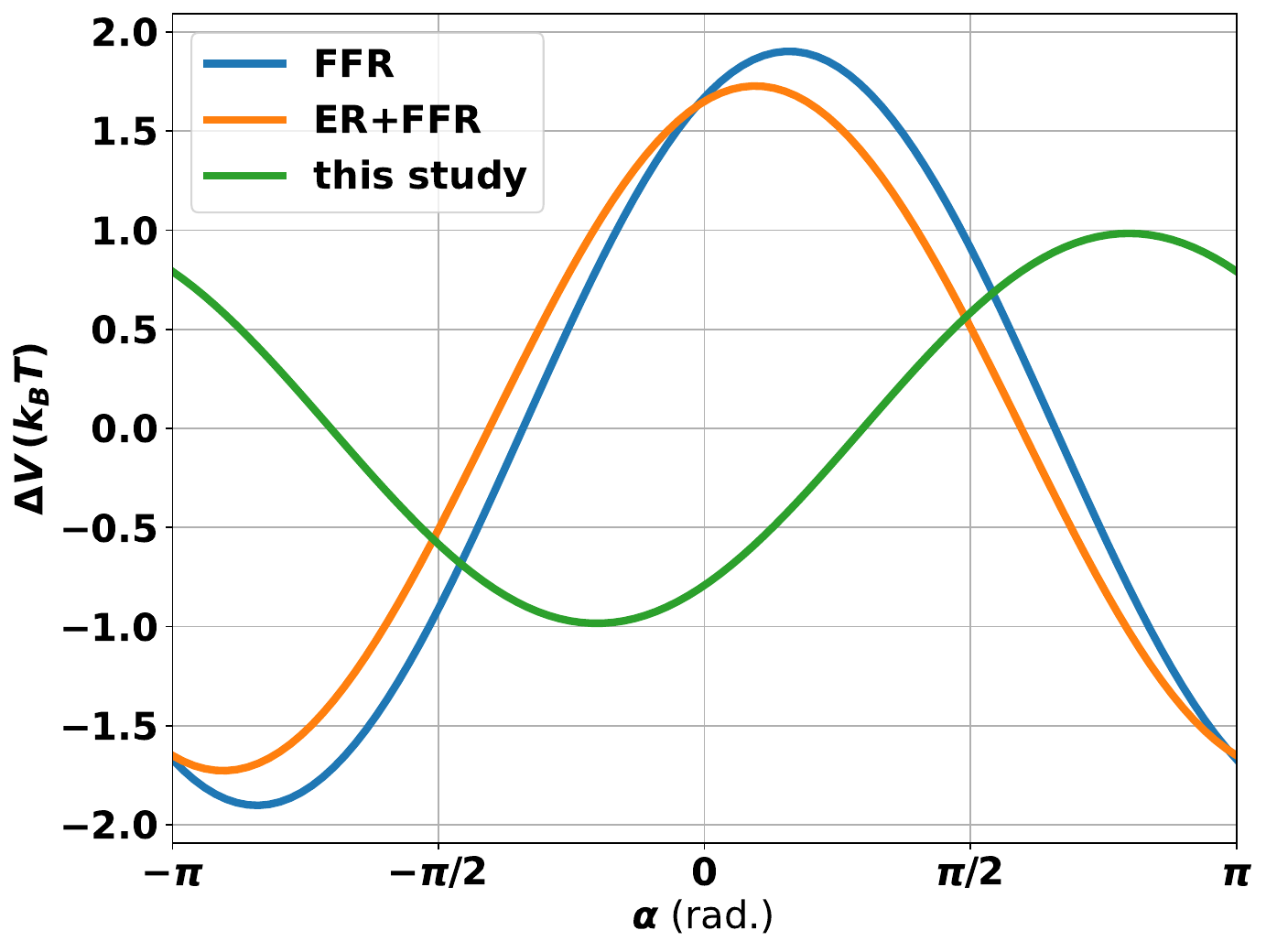}

    \end{subfigure}
    \begin{subfigure}[b]{0.5\textwidth}  %
        \caption[]{histogram of $\alpha$ dihedral angles}%
        \includegraphics[width=0.95\textwidth]{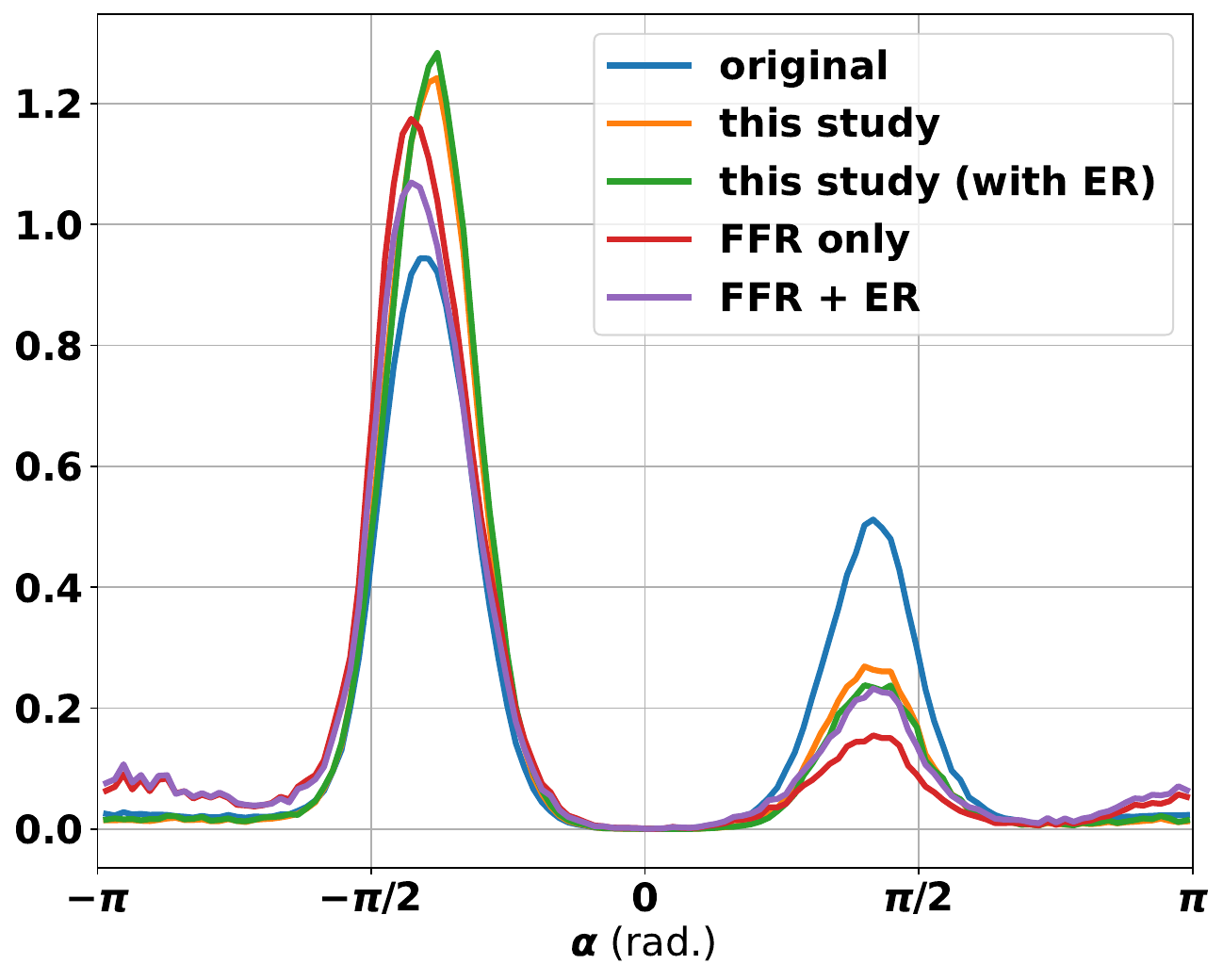}

    \end{subfigure}
    \caption
    {\small a) Force-field correction on $\alpha$ dihedral angles; comparison between this study (fully combined refinement) and previous\cite{gilardoni2024boosting} one (force-field refinement FFR only or combined with ensemble refinement FFR+ER). b) Average histogram of $\alpha$ dihedral angles; comparison between original one, this study (both reweighting with force-field correction only or including also contribution from ER) and previous study\cite{gilardoni2024boosting} (corresponding to the force-field corrections in Fig. \ref{fig:ff_refinement_fully}a).}
    \label{fig:ff_refinement_fully}
\end{figure}

Then, we search for the optimal hyperparameters by using the \texttt{hyper\_minimizer} routine of \texttt{MDRefine}. We perform the same splitting of observables and frames in training/test set as in the previous cross validation (with 5 different choices) and then minimize the average $\chi^2$ on cross-validated observables. We get as optimal hyperparameters $\alpha=959.9,\,\beta=167.9,\,\gamma=0.42 \, Hz^{-2}$.
Importantly, a grid search in a three-dimensional hyperparameter space
would have been extremely costly. In our case, the minimization algorithm converges in only 44 function evaluations, and is able to reach a cross-validation error very close to the minimum value in less than 10 evaluations (see plot \ref{fig:compare_chi2} in \textit{Supplementary}).

The resulting refinement of the Karplus equations is reported in Fig. \ref{fig:Karplus_refinement}. It overlaps to that determined in \cite{frohlking2023simultaneous} very well, except for a slight deviation regarding sugar puckering angles close to $\pm\pi/2$ and $\pi$.
The resulting force-field corrections are
reported in Fig. \ref{fig:ff_refinement_fully}a.
Interestingly, the correction proposed here appears completely different
from the one suggested in
\cite{gilardoni2024boosting}.
We also compared the population of the corrected dihedral angle, averaged over all systems, using the different approaches (Fig.~\ref{fig:ff_refinement_fully}a).
In spite of the significant difference in the force-field corrections, the resulting populations are quite similar. This can be rationalized considering the
steric hindrance, which allows only some regions for the $\alpha$ dihedral angles to be populated.
In other words, even though the corrections represented in Fig.~\ref{fig:ff_refinement_fully}a appears different, their effect on the sampled regions of Fig.~\ref{fig:ff_refinement_fully}b is comparable.
Indeed, by comparing the histograms in Fig. \ref{fig:ff_refinement_fully}b, we conclude that this refinement shifts the population of $\alpha$ dihedral angles in the same direction as the one determined in the previous work \cite{gilardoni2024boosting} through force-field refinement only or combined with ensemble refinement.
It is however more conservative due to the combined refinement of the forward model. Further, by comparing the two relative entropies $S[P_\phi|P_0]$ and $S[P|P_\phi]$ (Fig. \ref{fig:compare_rel_entropies} in Supplementary), we argue the contribution of flexible ensemble refinement does not further modify the ensembles in a significant way. This means that the force-field correction alone is able to capture a significant fraction of the required correction.

It is also possible to obtain an estimation of the performance of the different approaches on new observables using the $\chi^2$ evaluated on the cross-validation set. 
This is reported in Fig.~\ref{fig:compare_chi2}.
Here it can be seen that 
the combined refinement is expected to perform significantly better than the approach of Ref.~\cite{frohlking2023simultaneous}, in the sense that the $\chi^2$ on cross-validation observables is much lower.
Including an explicit refinement on a carefully-chosen force-field correction allows for suggesting a direction in which modifications to the structural ensembles improve the agreement with experimental data.

\section*{Discussion}

The integration of MD-generated trajectories with experimental measures is 
useful to both improve the accuracy of MD simulations and to provide a microscopic interpretation to experimental data.
Specifically, ensemble refinement methods can be used to minimally modify
MD-generated ensembles to match experiments that report averaged information.
Several variants of this idea have been proposed and used in the past years, based on the
idea of refining either ensembles, force-field parameters, or forward models
used to back-calculate experimental data.
However, a general tool to perform these types of analysis is missing.
We here developed the Python package \texttt{MDRefine}, which can be downloaded and installed from \url{https://pypi.org/project/MDRefine} and provides
a flexible implementation of the methods above.
\texttt{MDRefine} consists of a number of functions, starting from standard calculations (relative entropy, $\chi^2$) to more elaborate ones (minimization of the loss function and automatic search of optimal hyperparameters). It has been developed with a modular structure: the full input data defines a dictionary whose items are class instances, each of them containing all the information about a specific molecule (or a thermodynamic cycle for refinements with alchemical calculations). \texttt{MDRefine} is fairly customizable: the user can define functional forms for the force-field corrections and/or the forward models which are best suitable for his own cases of interest. The output is then returned as a folder with several files including optimal hyperparameters and coefficients, $\chi^2$ and relative entropies, Numpy arrays for the refined weights, and other relevant quantities.

We showed how to use \texttt{MDRefine} to reproduce data from recently published works \cite{piomponi2022molecular,frohlking2023simultaneous,gilardoni2024boosting}, where analyses were performed using custom scripts that are more difficult to adapt to new systems. These examples include fitting of partial charges of a modified nucleobase using alchemical simulations and fitting structural ensembles
of RNA oligomers using various combinations of force-field refinement, ensemble refinement, and forward-model refinement.
Jupyter notebooks that can be used to reproduce these analyses and can be used as starting points to analyze new systems are available on GitHub \url{https://github.com/bussilab/MDRefine}.

In addition to reproducing existing
analyses, we generalized them taking advantage of the flexibility of this new implementation. Specifically, we compare the effect of different regularization strategies, and introduce the idea of tuning the hyperparameters
by directly minimizing the cross-validation error. We consider this to be
especially valuable in applications where the number of hyperparameters is larger than two and an exhausive search is unfeasible.

It is important to recall that \texttt{MDRefine} is only designed to analyse and reweight existing trajectories, and not to generate new ones. This results in an important assumption, namely
that the modifications required to match experimental measurements are small refinements
\cite{rangan2018determination}. This holds whenever the original ensemble has a significant overlap with the experiments. 
For instance, in the case of force-field refinement, as soon as the deviations become higher than a certain threshold (fixed for example by the relative entropy or the effective sample size), it is advisable to perform a new simulation with the updated parameters\cite{norgaard2008experimental,cesari2019fitting,thaler2021learning,giorgetti2014predictive,tiana2016structural}.
Also ensemble refinement strategies based on the maximum entropy principle can be
iterated in this way, by running new simulations where linear restraints
are used as biases \cite{zhang2015topology,zhang2016shape}.
Parameters might also be improved on-the-fly, resulting in a much larger
statistical significance, at the price of a more complex implementation \cite{cesari2016combining, white2015designing, marinelli2015ensemble, gil2016empirical, white2014efficient, hocky2017coarse}. Finally, replica-based methods where the time average is replaced with a replica average, might also be used \cite{lindorff2005simultaneous, cavalli2013molecular, roux2013statistical, bonomi2016metainference}.

\texttt{MDRefine} provides an off-the-shelf implementation of complex state-of-the-art algorithms which allow to refine
structural ensembles for complex molecular systems. We hope that proving an open source and well-documented tool
will facilitate the use of these algorithms in the molecular dynamics community.

\section*{Funding}
V.P. was supported by the European Union---NextGenerationEU within the project PNRR (PRP@CERIC)

\bibliography{Main}  

\newpage
\section*{Supplementary}

\subsection*{S1 - Computation of derivatives}

Let's remind $\mathcal L_1$, the loss function with the explicit dependence $\lambda^*(\mu)$
\begin{equation}
\mathcal{L}_1(\mu) = \mathcal{L}(\lambda^*(\mu),\mu)=-\alpha\Gamma(\lambda^*(\mu);\mu)+\beta R_1(\phi)+\gamma R_2(\theta)
\end{equation}
where we introduced the shortcut $\mu=(\phi,\theta)$. The optimal parameters are

\begin{equation}
\mu^* = \arg\min_{\mu}\Bigl (-\alpha\min_\lambda\Gamma(\lambda;\mu)+\beta R_1(\phi)+\gamma R_2(\theta)\Bigr);
\end{equation}

\begin{equation}
\lambda^* = \arg\min_\lambda\Gamma(\lambda;\mu^*).
\end{equation}

In order to search for the optimal coefficients $\lambda^*,\mu^*$, we rely on the BFGS algorithm (or L-BFGS-B if there are boundaries on $\lambda$, due to $\min,\max$ in the $\chi^2$, like for \textit{unobserved NOEs} data) implemented by \texttt{Scipy}. This requires the computation of the derivatives, which we can carry out by leveraging the automatic differentiation performed by \texttt{JAX}. However, $\mathcal L_1$ depends on $\mu$ also through $\lambda^*$, namely through the full minimization algorithm employed for $\Gamma$. To overcome this issue, because $\lambda^*$ is a point of minimum for $\mathcal L$, the derivative simplifies to
\begin{equation}
\frac{\partial \mathcal{L}_1(\mu)}{\partial\mu_j} = \frac{\partial\mathcal{L}(\lambda,\mu)}{\partial\lambda_k}\Big|_{\lambda^*(\mu)} \frac{\partial \lambda_k^*(\mu)}{\partial\mu_j} + \frac{\partial \mathcal L(\lambda,\mu)}{\partial\mu_j}\Big|_{\lambda^*(\mu)} = \frac{\partial \mathcal L(\lambda,\mu)}{\partial\mu_j}\Big|_{\lambda^*(\mu)}
\end{equation}
and so it can be implemented via automatic differentiation of $\texttt{loss\_function}$ with given $\lambda^*$ passed as $\texttt{fixed\_lambdas}$.

Our aim is to compute how the $\chi^2$ in validation, evaluated at the optimal parameters $\lambda^*$ and $\mu^*$ found in training, varies with the hyperparameters $\alpha,\beta,\gamma$, and to compute its derivatives. By employing the chain rule, this task can be split into two parts: compute separately the derivatives of the $\chi^2$ w.r.t. the optimal coefficients $\lambda^*,\mu^*$ and those of the optimal coefficients $\lambda^*,\mu^*$ w.r.t. the hyperparameters, then multiply opportunely the derivatives. For the sake of clarity, in the following we will use the Einstein notation (sum over repeated index is implicit). 

Let's focus firstly on the second part and let's make the dependence on the hyperparameters explicit: our aim is to compute the derivatives of the optimal parameters $\lambda^*$ and $\mu^*$ w.r.t. the hyperparameters $\alpha,\beta,\gamma$. The optimal coefficients are determined by the equations
\begin{equation}
\lambda^*(\mu;\alpha) = \arg\min_\lambda \Gamma(\lambda;\mu,\alpha);
\end{equation}
\begin{equation}
\mu^*(\alpha,\beta,\gamma)=\arg\min_\mu \mathcal L_1 (\mu;\alpha,\beta,\gamma).
\end{equation}

In absence of boundaries, the optimal parameters are interior to the domain of the loss function. In such points the gradient is zero:
\begin{equation}
\frac{\partial\Gamma(\lambda;\mu,\alpha)}{\partial\lambda}\Big|_{\lambda^*} = 0\,\,\,\mathrm{and}\,\,\,
\frac{\partial \mathcal L_1(\mu;\alpha,\beta,\gamma)}{\partial\mu}\Big|_{\mu^*}=0.
\end{equation}
Hence, we can exploit the implicit function theorem and get to the following relations:
\begin{equation}
\frac{\partial\lambda_i^*}{\partial(\mu_j,\alpha)} = -(H_{in}^{-1})_{ik} \frac{\partial^2 \Gamma(\lambda;\mu,\alpha)}{\partial \lambda_k \partial(\mu_j,\alpha)}
\end{equation}

\begin{equation}
\frac{\partial\mu_i^*}{\partial(\alpha,\beta,\gamma)} = -(H^{-1})_{ij}\frac{\partial^2 \mathcal L_1(\mu;\alpha,\beta,\gamma)}{\partial\mu_j \partial(\alpha,\beta,\gamma)}
\end{equation}
where $H_{in}$ is the Hessian of the internal minimization (over $\lambda$)

\begin{equation}
H_{in\, ab} = \frac{\partial^2 \Gamma(\lambda;\mu,\alpha)}{\partial \lambda_a \partial\lambda_b}
\end{equation}
and $H$ is the Hessian of the minimization over $\mu$

\begin{equation}
H_{ab} = \frac{\partial^2 \mathcal L_1(\mu;\alpha,\beta,\gamma)}{\partial \mu_a \partial\mu_b}.
\end{equation}

The internal Hessian $H_{in}$ can be directly computed through automatic differentiation of $\Gamma(\lambda;\mu,\alpha)$ by JAX (which can be equivalently done for \texttt{gamma\_function} or for \texttt{loss\_function}$/\alpha$ with \texttt{fixed\_lambdas} not \texttt{None}), while for the second Hessian $H$ a further calculation is required. Indeed, we have to consider also the dependence of $\mathcal L_1$ on $\mu$ through $\lambda^*$. 
Consequenly, the Hessian results

\begin{equation}
\begin{split}
\frac{\partial^2\mathcal L_1}{\partial\mu_j \partial\mu_k} &
= \Bigl[\frac{\partial}{\partial\lambda_l}\Bigl(\frac{\partial\mathcal L}{\partial\mu_j}\Bigr)\Bigr]_{\lambda^*} \frac{\partial \lambda_l^*}{\partial\mu_k} + \frac{\partial^2\mathcal L}{\partial\mu_j\mu_k} \\
&= +\alpha \frac{\partial^2 \Gamma}{\partial\mu_j\partial\lambda_l}
(H_{in}^{-1})_{lm}\frac{\partial^2\Gamma}{\partial\lambda_m\partial\mu_k} + \frac{\partial^2\mathcal L}{\partial\mu_j\partial\mu_k}
\end{split}
\end{equation}
where the first term has to be summed over the
molecular systems used in cross validation, which are possibly
different from those used in training.
Similarly, it holds
\begin{equation}
\frac{\partial^2\mathcal L_1}{\partial\mu_j\partial\alpha} = \frac{\partial}{\partial\alpha} \frac{\partial\mathcal L(\lambda^*(\mu,\alpha),\mu;\alpha,\beta,\gamma)}{\partial\mu_j}
= \frac{\partial^2 \mathcal L}{\partial\mu_j \partial\lambda_k}\frac{\partial\lambda_k}{\partial\alpha}+\frac{\partial^2\mathcal L}{\partial\mu_j\partial\alpha}.
\end{equation}
Also these terms can be computed through automatic differentiation.

Then, let's focus on the first part, namely, the derivatives of the $\chi^2$ with respect to the optimal parameters $\lambda^*,\mu^*$. The full dependence is through the average value of the validating observables $\tilde g_i(\theta^*)$ (computed with the optimized forward model $\theta^*$) on the ensemble corresponding to the optimal parameters $\lambda^*,\mu^*$:
\begin{equation}
\langle \tilde g_i(\theta^*(\alpha,\beta,\gamma))\rangle_{\lambda^*(\mu^*(\alpha,\beta,\gamma);\alpha),\,\mu^*(\alpha,\beta,\gamma)}.
\end{equation}
These derivatives can be directly computed through automatic differentiation. Let's conclude our calculation by applying the chain rule:
\begin{equation}
\frac{\partial\chi^2}{\partial\alpha}
= \frac{\partial\chi^2}{\partial\mu_i}\frac{\partial\mu_i^*}{\partial\alpha} + \frac{\partial\chi^2}{\partial\lambda_i}\Bigl(\frac{\partial\lambda_i^*}{\partial\mu_j}\frac{\partial\mu_j^*}{\partial\alpha}+\frac{\partial\lambda_i^*}{\partial\alpha}\Bigr)
\end{equation}

\begin{equation}
\frac{\partial\chi^2}{\partial(\beta,\gamma)}
= \frac{\partial\chi^2}{\partial\mu_i}\frac{\partial\mu_i^*}{\partial(\beta,\gamma)} + \frac{\partial\chi^2}{\partial\lambda_i}\frac{\partial\lambda_i^*}{\partial\mu_j}\frac{\partial\mu_j^*}{\partial(\beta,\gamma)}.
\end{equation}

Now, each term on the right side of the equations can be computed through automatic differentiation and we just have to multiply them as indicated in order to get the derivatives of $\chi^2$ w.r.t. the hyperparameters $\alpha,\beta,\gamma$. This will facilitate the search for the optimal hyperparameters, without the need for a scan in 3 dimensions, which may be cumbersome. Lastly, the minimization can be sped up
considering the fact that one needs
to explore several orders of magnitude in the hyperparameter space.
It is then convenient to change variable and perform the minimization
considering as hyperparameters $\log \alpha$, $\log \beta$, and $\log \gamma$, using once again the chain rule for derivatives
$\frac{\partial}{\partial\log\alpha} = \alpha\frac{\partial}{\partial\alpha}$, and analogous expressions for the other hyperparameters.

These calculations can be adapted to the presence of boundaries for the $\lambda$ parameters (like in the case of \textit{unobserved NOEs}). In these cases, when the optimal solution is on the boundary $\lambda_j = 0$ for some $j$, the derivatives of these $\lambda_j$ parameters w.r.t. $\alpha,\mu$ is zero. To prove this statement, let's consider for simplicity the case with the constraint $\lambda_j > 0$ for some $j$. When the optimal solution is on the boundary $\lambda_j = 0$ it means that, if we could remove the boundary, the minimum of the loss function $\Gamma(\lambda;\alpha,\mu)$ would be inner to the forbidden region $\lambda_j < 0$. Hence, by continuity of the $\Gamma$ function with respect to $\alpha$ (such as for $\mu$), when varying $\alpha$ (or $\mu$) by an arbitrarily small value, the minimum will still be in the forbidden region and the value of $\lambda_j$ will still be zero, without changing. This holds for every function $f(\lambda;\alpha,\mu)$ which is continuous in its arguments, independently of its convexity or not.
Notice that since the boundaries are in the form of linear inequalities on $\lambda$, the function $\Gamma$ is expected to be convex also on the boundary.

\subsection*{S2 - Implementation of \texttt{minimizer}}

To determine the optimal $\lambda^*(\phi,\theta)$ at given $(\phi,\theta)$ we can employ the same \texttt{gamma\_function} previously described. Notice \texttt{gamma\_function} does not include either the computation of the observables through the forward model $\theta$ or the reweighting through the force-field correction $\phi$ (namely, the dependence of $\Gamma$ on $\theta,\phi$), rather they should be externally computed and then passed to \texttt{gamma\_function} as input observables and weights.

Regarding the full loss function, we implemented $\mathcal L_1(\phi,\theta)$ in \texttt{loss\_function} by inserting the search for optimal $\lambda^*(\phi,\theta)$. Its inputs are the Numpy array of $(\phi,\theta)$ coefficients, the \texttt{data} object, the \texttt{regularization} dictionary, the hyperparameters $\alpha,\beta,\gamma$ (by default $+\infty$, i.e., no refinement in that direction) and some other variables. The \texttt{regularization} dictionary has the two keys \texttt{force\_field\_reg}, \texttt{forward\_model\_reg}; the first item can be either a string (\texttt{KL divergence}, \texttt{plain l2}) or a user-defined function acting on $\phi$; the second item is a user-defined function acting on $\theta$ and $\theta_0$ (for example, an $L^2$ regularization).

The other inputs include: \texttt{fixed\_lambdas} (by default \texttt{None}), \texttt{gtol\_inn} and \texttt{bounds} (tolerance and boundaries for the inner minimization of $\lambda$; default values $10^{-3}$ and \texttt{None}, respectively), \texttt{if\_save} (Boolean variable, by default \texttt{False}). For \texttt{if\_save = True}, \texttt{loss\_function} does not return just its numerical value, rather an object with all the quantities appearing in Eq. \ref{eqn:loss_complete}, such as the parameters $(\lambda,\phi,\theta)$ (when defined), the new weights, the $\chi^2$, the regularization terms.

Whenever the input variable \texttt{fixed\_lambdas} is not \texttt{None}, \texttt{loss\_function} implements the computation of $\mathcal L_1(\phi,\theta)$ through substitution of the given $\lambda$ into $\Gamma(\lambda;\phi,\theta)$ in place of the inner minimization of $\Gamma$ over $\lambda$ (the two computations are the same if $\alpha$ is infinite, namely, no ensemble refinement). This distinction turns out to be crucial when computing the gradient function $\frac{\partial\mathcal L_1(\phi,\theta)}{\partial(\phi,\theta)}$ \texttt{gradient\_fun} with automatic differentiation (see \textit{Supplementary - Computation of derivatives}).

The \texttt{loss\_function} and its \texttt{gradient\_fun} are evaluated together inside \texttt{loss\_function\_and\_grad}, which is employed in \texttt{minimizer} in order to get the optimal solution. In this way, passing to \texttt{minimizer} the \texttt{data} object, the \texttt{regularization} dictionary and the hyperparameters (by default $\alpha,\beta,\gamma=+\infty$, i.e., no refinement in those directions), we get an output object \texttt{Result} containing the optimal solution $(P, \phi, \theta)$ and its properties.

\subsection*{S3 - Kish sample size and entropy}

Let's consider a sample $\{ x_1, ..., x_n\}$ (assumed to be uncorrelated) with corresponding normalized weights $\{w_1, ..., w_n\}$. If the weights are uniform, then the effective sample size is simply given by the number of points $n_{eff}=n$. Otherwise, we can obtain a good estimation of the effective sample size by
using the Kish sample size, defined as
the inverse of the average weight
\begin{equation}
n_{eff}=K[\{w_i\}]\coloneq\langle w\rangle_w^{-1}=\Bigl(\sum_i w_i^2\Bigr)^{-1},
\label{eqn:Kish_size}
\end{equation}
This definition of $n_{eff}$ reduces to $n_{eff}=n$ if each sample contributes with uniform weight, while it returns the number of ``dominant" frames if some of them count much more than the others. For example, if we have a non-normalized set of weights $\{1,...,1,N,...,N\}$ with weights $1$ and $N$ repeated $n-m$ and $m$ times, respectively, then the effective sample size is $n_{eff}\simeq m$ if $n-m\ll Nm$ (the subset of small weights $w_i = 1$ dominates over the other) and $n_{eff}\simeq n-m$ in the opposite limit. Clearly, it is always $n_{eff}\leq n$ and the equality holds only for uniform weights; it is also $n_{eff}>1$ and small values are achieved when one frame contributes much more than the others.  

The relation with the entropy $S$ is straightforward. It suffices to change the weighted arithmetic average $\langle...\rangle$ to the weighted geometric one in Eq. \ref{eqn:Kish_size} to get

\begin{equation}
e^{S[\{w_i\}]} = \Bigl(\prod_i w_i^{w_i}\Bigr)^{-1},
\end{equation}
which is always in between $1<e^{S[\{w_i\}]}\leq n$. Hence, the effective number of frames can be estimated also through $e^{S[\{w_i\}]}$, in a way which could be less sensitive to
the fact that weights could vary over 
orders of magnitude.
Moreover, by the Jensen's inequality, it is always $e^{S[\{w_i\}]}\geq K[\{w_i\}]$.

Then, the relative entropy $S_{rel}$ can be used to quantify the discrepancy of the distribution $P$, represented by the set of weights $\{w_i\}$, from $P_0$, represented by $\{w_{0i}\}$: 
\begin{equation}
e^{S_{rel}[P|P_0]} = \Bigl[\prod_i \Bigl(\frac{w_i}{w_{0i}}\Bigr)^{w_i}\Bigr]^{-1}.
\end{equation}

The relative entropy is always $S_{rel}\leq 0$ but can be very large in absolute value for $P$ very different from $P_0$, with a lower bound determined by the number of frames. Notice it is the geometric average (over $w$) of the ratio $w/w_0$; changing it to the arithmetic average we get the relative Kish size\cite{frohlking2022automatic}
\begin{equation}
K_{rel}[P|P_0] = \Bigl\langle\frac{w}{w_0}\Bigr\rangle_w^{-1}.
\label{eqn:rel_kish_size}
\end{equation}

If the reference ensemble is uniform ($w_0=\frac{1}{n}$ constant), then the relative Kish size (and analogously $e^{S_{rel}[P|P_0]}$) reduces to the ratio $n_{eff}/n$.

Finally, the Kish size ratio (KSR) for an ensemble $P$ reweighted from $P_0$ has been defined\cite{piomponi2022molecular} as the ratio between the two Kish sample sizes of $P$ and $P_0$, respectively:
\begin{equation}
KSR = \frac{\langle w\rangle_w^{-1}}{\langle w_0 \rangle_{w_0}^{-1}}.
\end{equation}

By following the analogy between Kish size and entropy,
which share the same definition but using arithmetic and geometric averages, respectively, the Kish size ratio can be
considered analogous to the exponential of the entropy difference
$e^{S[P]-S[P_0]}$.

\subsection*{S4 - Force-field refinement with alchemical calculations}

\begin{figure}
    \captionsetup{justification=justified, singlelinecheck=off}
    \begin{subfigure}[b]{0.45\textwidth} %
        \caption[]{min. loss function for $L^2$ regularization; the overfitting transition is quite evident}%
        \includegraphics[width=\textwidth]{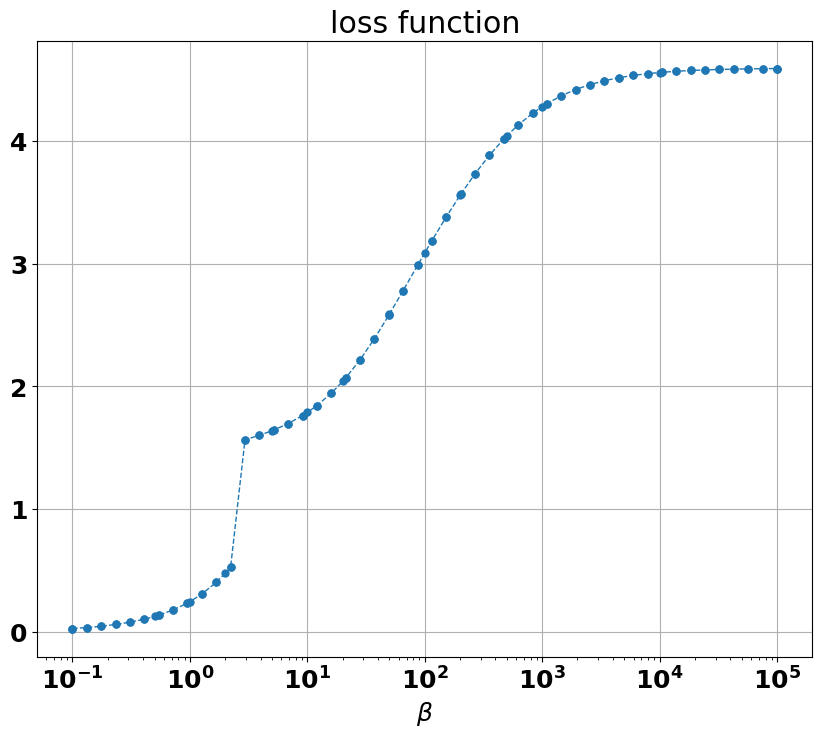}

    \end{subfigure}
    \begin{subfigure}[b]{0.47\textwidth}  %
        \caption[]{min. loss function for relative entropy regularization}%
        \includegraphics[width=\textwidth]{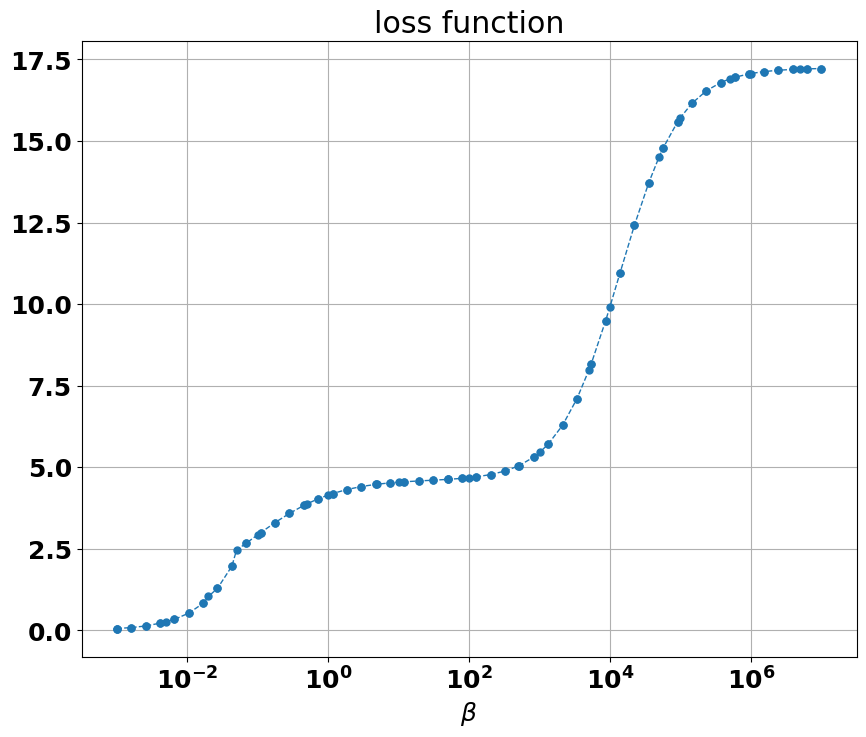}

    \end{subfigure}
    \vskip\baselineskip
    \begin{subfigure}[b]{0.49\textwidth}   %
        \caption[]{$\chi^2$ for different thermodynamic cycles and average KSR for $L^2$ regularization}%
        \includegraphics[width=\textwidth]{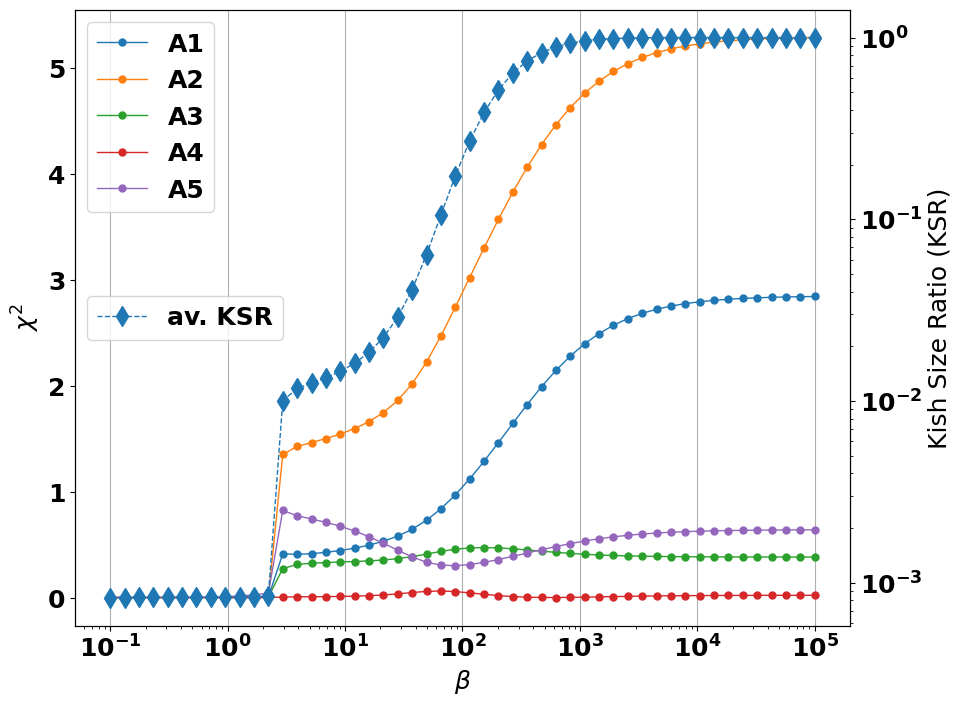}

    \end{subfigure}
    \begin{subfigure}[b]{0.5\textwidth}   %
        \caption[]{$\chi^2$ for different thermodynamic cycles and average KSR for relative entropy regularization}%
        \includegraphics[width=\textwidth]{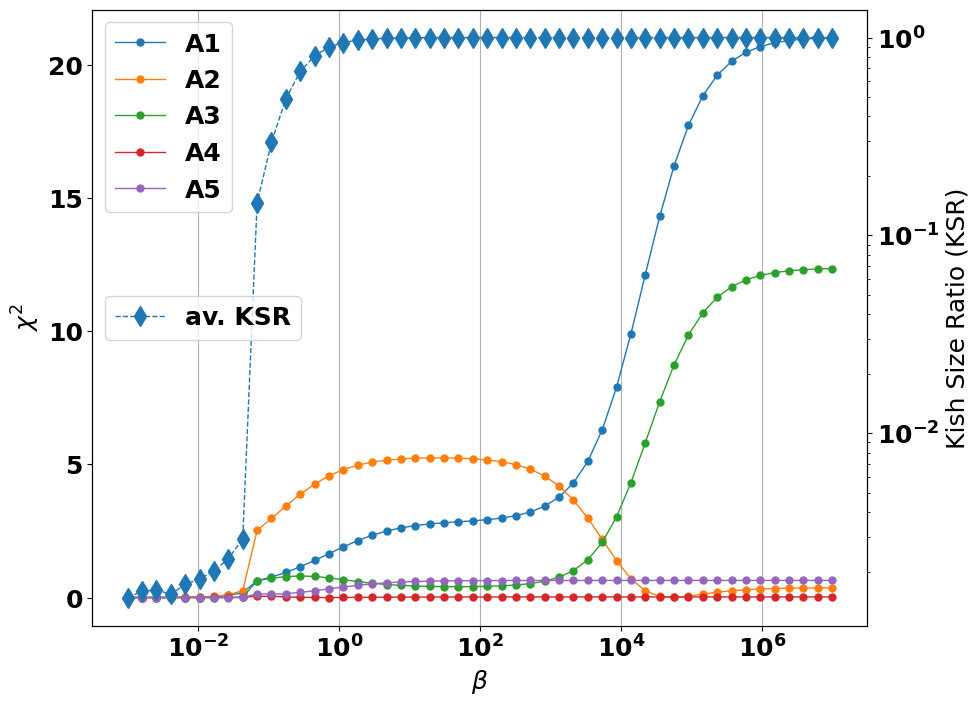}
        
    \end{subfigure}
    \caption[ alchemical ]
    {\small Loss function, average Kish size ratio and $\chi^2$ for each thermodynamic cycle at the optimal solution found at given hyperparameter $\beta$ and regularization ($L^2$ on charges or relative entropy). The overfitting transition is quite evident in the case of $L^2$ regularization, while it is smoother for the relative entropy regularization.}
    \label{fig:alchemical_refinement_appendix}
\end{figure}

In Fig. \ref{fig:alchemical_refinement_appendix} we report the minimum value of the loss function, the $\chi^2$ for the different thermodynamic cycles and the average Kish size ratio (KSR) as a function of the hyperparameter $\beta$. The overfitting transition is quite evident in the case with $L^2$ regularization, while it is smoother in the second case. For the sake of clarity, in section S2, we report the definitions of Kish size ratio and relative Kish size, introduced in previous works \cite{piomponi2022molecular, frohlking2023simultaneous}, and highlight their relationships with the entropy or the relative entropy.

\subsection*{S5 - Forward-model refinement + max. entropy}

\begin{figure}
    \centering
    \includegraphics[width=1\linewidth]{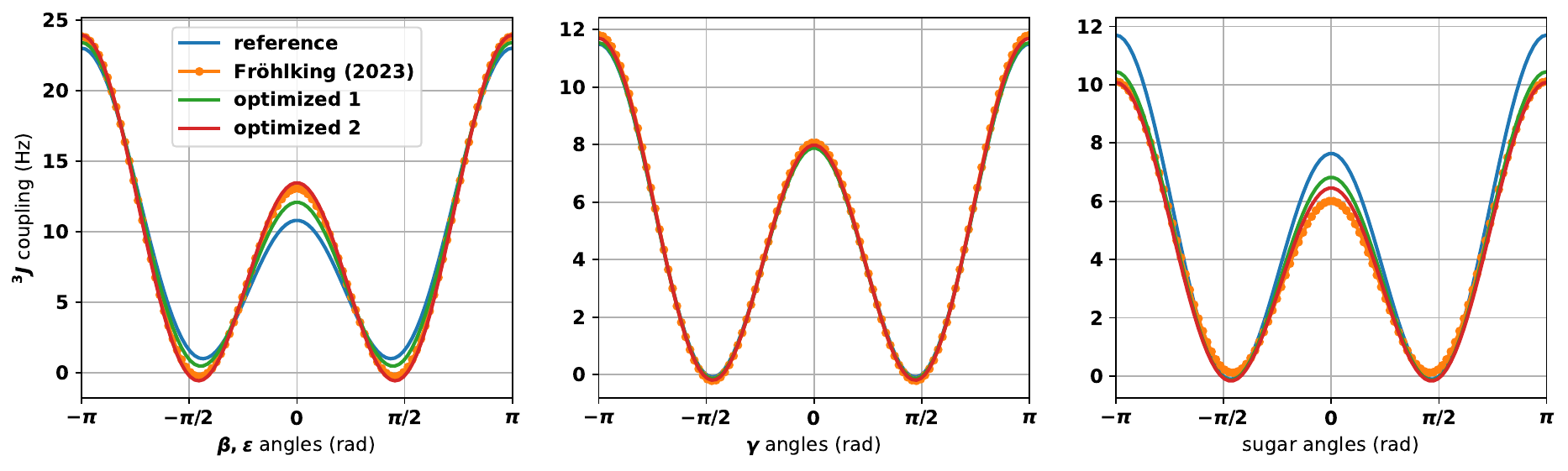}
    \caption{Refinement of the empirical coefficients for the Karplus equations with Ensemble + Forward Model Refinement; comparison between \textit{optimized 1} and \textit{optimized 2} (see Table \ref{tab:karplus}.}\label{fig:Karplus_refinement_compare}
\end{figure}

In Table \ref{tab:karplus} we report the optimized empirical coefficients of the Karplus equations, comparing them with values already known in the literature. In Fig. \ref{fig:Karplus_refinement_compare} we report the 

\begin{figure}
    \centering
    \includegraphics[width=0.8\linewidth]{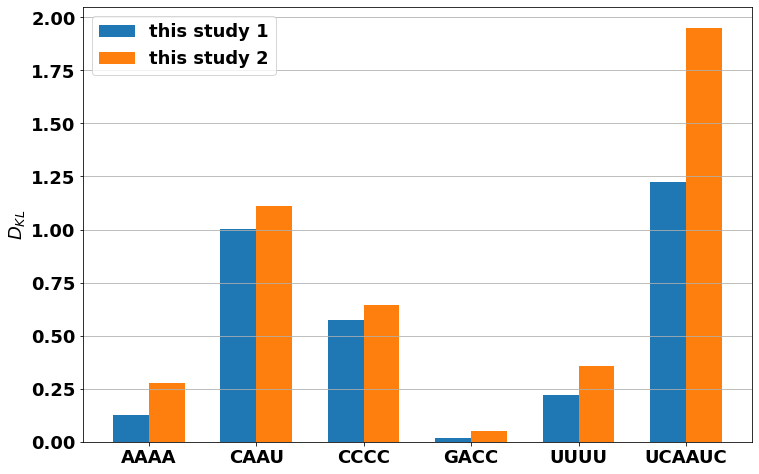}
    \caption{Refinement of Karplus equations: Kullback-Leibler divergences of the refined ensembles with respect to the original ones (optimized 1 with same hyperparameters as in \cite{frohlking2023simultaneous}; 2 with optimized hyperparameters).}
    \label{fig:Karplus_refinement_DKL}
\end{figure}

\begin{figure}
    \centering
    \includegraphics[width=1\linewidth]{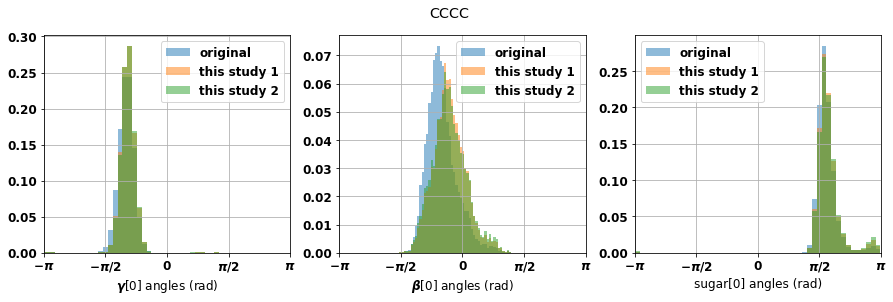}
    \includegraphics[width=1\linewidth]{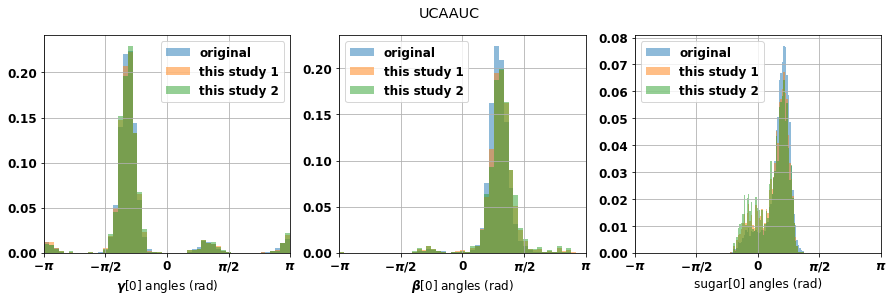}

    \caption{Weighted histograms of some dihedral angles of CCCC and UCAAUC oligomers, as examples (comparison between original ensemble and reweighted ones).}
    \label{fig:Karplus_refinement_hist}
\end{figure}

In Fig. \ref{fig:Karplus_refinement_DKL} we report the Kullback-Leibler divergence of the refined ensembles with respect to the original ones, represented by MD simulations. These values are quite small: the ensembles do not change significantly; this is confirmed by Fig. \ref{fig:Karplus_refinement_hist} showing weighted histograms of some dihedral angles.

\begin{table}
\captionsetup{justification=justified, singlelinecheck=off} %
\begin{tabular}{c|c|c|c}
name & A & B & C \\ %

\hline
\multicolumn{4}{c}{$\beta,\,\varepsilon$} \\
\hline
reference & 15.3 & -6.1 & 1.6 \\
Fröhlking & 18.34 & -5.39 & 0.11 \\
optimized 1 & 16.80 & -5.66 & 0.94 \\
optimized 2 & 18.91 & -5.23 & -0.21 \\
fully combined & 18.13 & -5.29 & 0.072 \\

\hline
\multicolumn{4}{c}{$\gamma$} \\
\hline
reference & 9.7 & -1.8 & 0 \\
Fröhlking & 10.07 & -1.87 & -0.13 \\
optimized 1 & 9.77 & -1.83 & -0.057 \\
optimized 2 & 9.95 & -1.86 & -0.11 \\
fully combined & 9.92 & -1.76 & -0.12 \\

\hline
\multicolumn{4}{c}{\textit{sugar}} \\
\hline
reference & 9.67 & -2.03 & 0 \\
Fröhlking & 7.81 & -2.05 & 0.25 \\
optimized 1 & 8.68 & -1.81 & -0.049 \\
optimized 2 & 8.34 & -1.81 & -0.069 \\
fully combined & 6.99 & -1.64 & 0.67 \\
\hline
\bottomrule
\end{tabular}
\caption{Empirical coefficients $A, B, C$ (in $Hz$) of the Karplus equations. Comparison between: reference values; coefficients determined by Fröhlking et al. in \cite{frohlking2023simultaneous} (hyperparameters $\alpha=174.33288,\,\gamma=6.86664/2 \, Hz^{-2}$ - factor $1/2$ because of different regularization prefactor); coefficients determined in this work (optimized 1, i.e., with same hyperparameters as in \cite{frohlking2023simultaneous}; optimized 2, i.e., with optimal hyperparameters found in this work $\alpha=46.31,\,\gamma=0.96 \, Hz^{-2}$); coefficients determined in this work with the fully combined approach ($\alpha=959.86,\,\beta=167.94,\,\gamma=0.42\,Hz^{-2}$).}
\label{tab:karplus}
\end{table}

\subsection*{S6 - Fully combined refinement}

\begin{figure}
    \centering
    \includegraphics[width=1\linewidth]{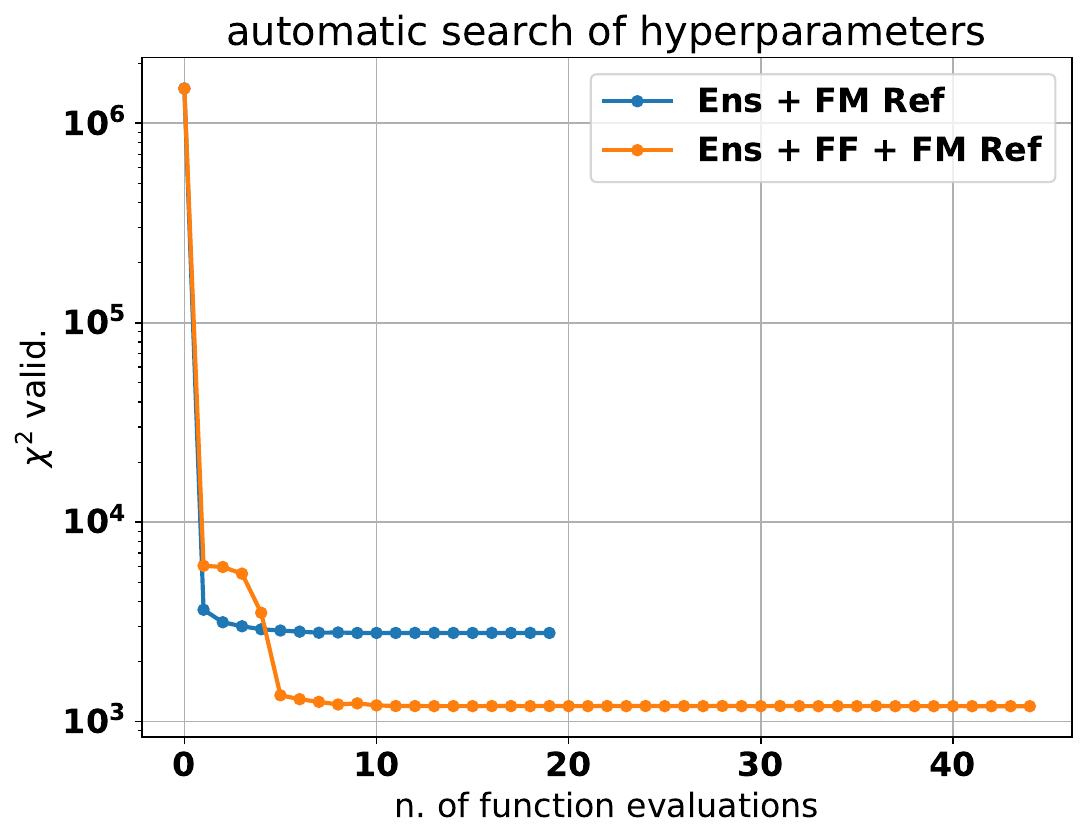}

    \caption{Average value of $\chi^2$ on validating observables in the course of the automatic search for the optimal hyperparameters. Comparison among Ens + FM (Ensemble + Forward Model) Refinement and Ens + FF + FM (the fully combined) Refinement. Starting from the initial values $\alpha=\beta=1,\,\gamma=1 \, Hz^{-2}$, we get the final values $\chi^2=2784.1$ for Ens+FM Ref. ($\alpha=46.31,\,\gamma=0.96\,Hz^{-2}$) and $\chi^2=1195.6$ for the fully combined approach ($\alpha=959.9,\,\beta=167.9,\,\gamma=0.42 \, Hz^{-2}$).}%
    \label{fig:compare_chi2}
\end{figure}

\begin{figure}
    \centering
    \includegraphics[width=0.8\linewidth]{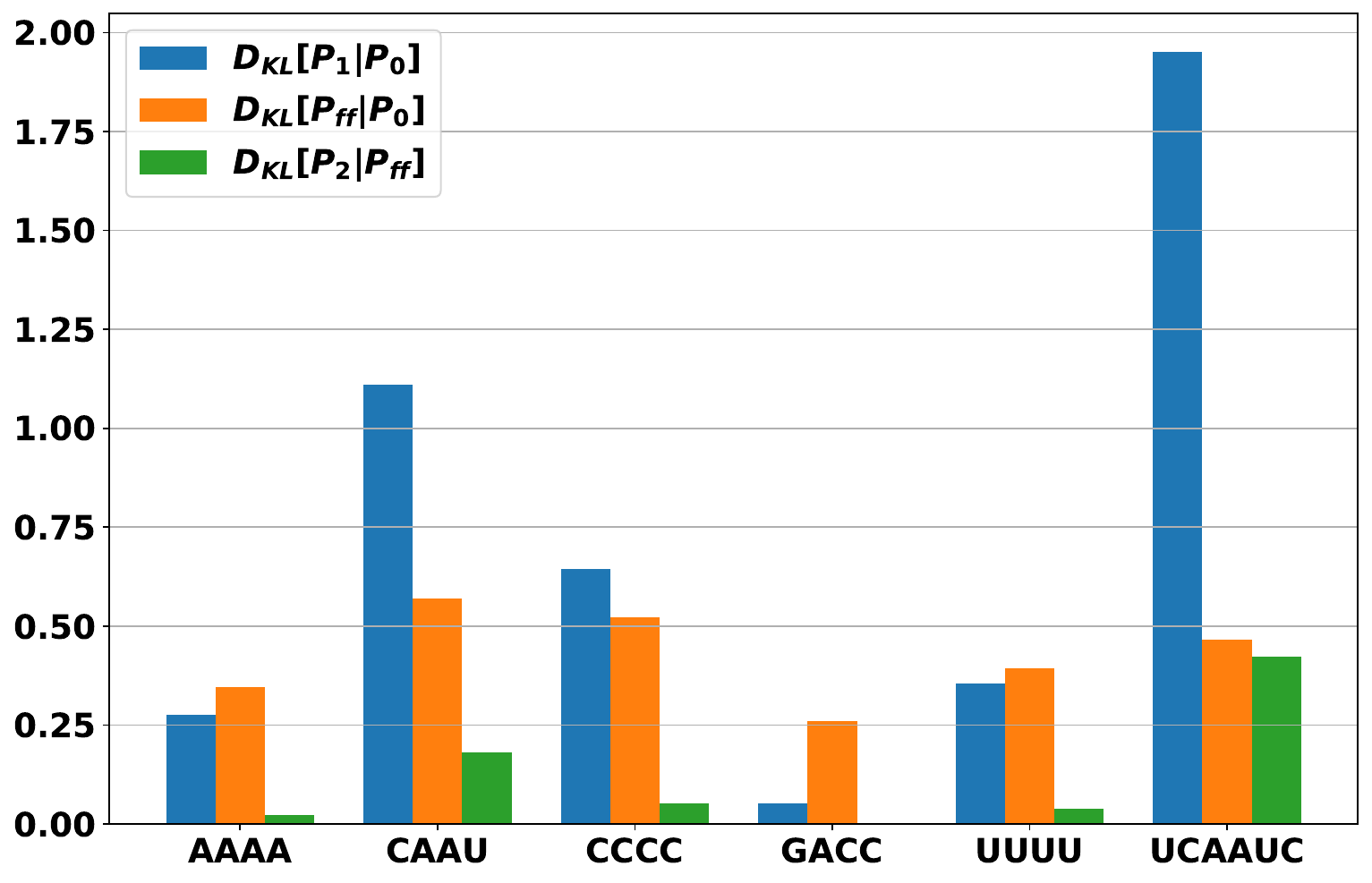}
    \caption{Comparison of the Kullback-Leibler divergences for different molecular systems when performing E+FM refinement ($P_1$) or the fully combined approach, with $P_{ff}$ the force-field correction ensemble and $P_2$ the final ensemble in the loss function (Eq. \ref{eqn:loss_complete}).}
    \label{fig:compare_rel_entropies}
\end{figure}

\begin{figure}
    \centering
    \includegraphics[width=0.8\linewidth]{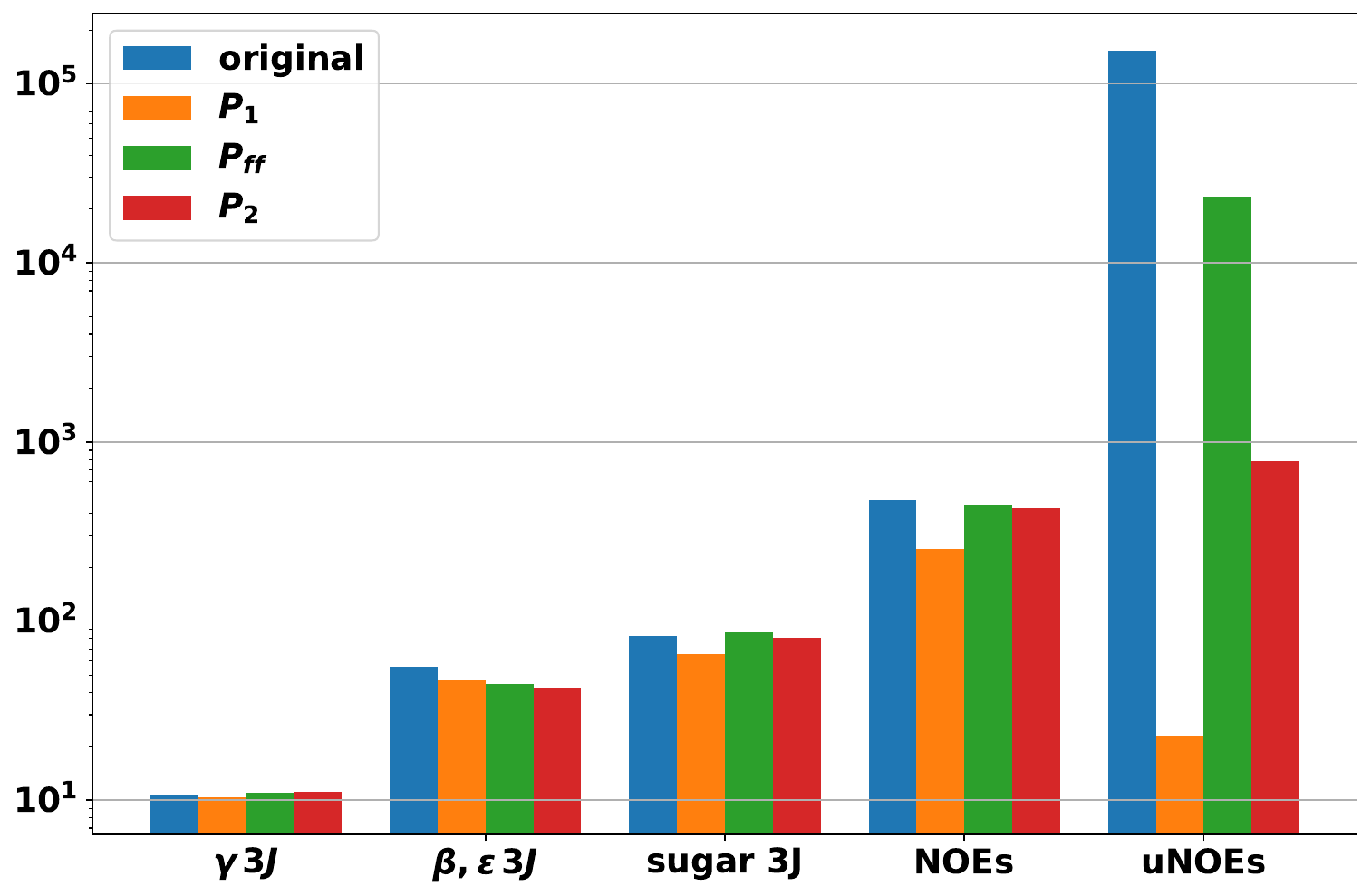}
    \caption{Comparison of the $\chi^2$ for different experimental measures when performing E+FM refinement ($P_1$) or the fully combined approach, with $P_{ff}$ the force-field correction ensemble and $P_2$ the final ensemble in the loss function (Eq. \ref{eqn:loss_complete}) (sum over different molecular systems).}
\end{figure}

\end{document}